\documentclass[preprint,superscriptaddress,amsmath,amssymb,prd,aps,showpacs,floatfix,nofootinbib]{revtex4-1}
\usepackage{graphicx}
\usepackage{dcolumn}
\usepackage{bm}
\usepackage{mathrsfs}
\usepackage{hyperref}
\usepackage{amsmath}
\usepackage{amssymb}
\usepackage{float}
\usepackage{dcolumn}
\usepackage{multirow}

\begin{document}
\title{Line shape and $D^{(\ast)}\bar D^{(\ast)}$ probabilities of $\psi(3770)$ from the $e^+e^-\to D\bar D$ reaction}
\date{\today}
\author{Q.~X.~Yu}
\email{qixinyu@ific.uv.es}\email{yuqx@mail.bnu.edu.cn}
\affiliation{College of Nuclear Science and Technology, Beijing Normal University, Beijing 100875, China}
\affiliation{Departamento de
F\'{\i}sica Te\'orica and IFIC, Centro Mixto Universidad de
Valencia-CSIC Institutos de Investigaci\'on de Paterna, Aptdo.22085,
46071 Valencia, Spain}

\author{W.~H.~Liang}
\email{liangwh@gxnu.edu.cn}
\affiliation{Department of Physics, Guangxi Normal University, Guilin 541004, China}
\affiliation{Guangxi Key Laboratory of Nuclear Physics and Technology, Guangxi Normal University, Guilin 541004, China}

\author{M.~Bayar}
\email{melahat.bayar@kocaeli.edu.tr}
\affiliation{Department of Physics, Kocaeli University, Izmit 41380, Turkey}

\author{E.~Oset}
\email{oset@ific.uv.es}
\affiliation{Departamento de
F\'{\i}sica Te\'orica and IFIC, Centro Mixto Universidad de
Valencia-CSIC Institutos de Investigaci\'on de Paterna, Aptdo.22085,
46071 Valencia, Spain}
\affiliation{Department of Physics, Guangxi Normal University, Guilin 541004, China}

\begin{abstract}
We have performed a calculation of the $D\bar D$, $D\bar D^\ast$, $D^\ast\bar D$, $D^\ast\bar D^\ast$ components in the wave function of the $\psi(3770)$. For this we make use of the $^3P_0$ model to find the coupling of $\psi(3770)$ to these components, that with an elaborate angular momentum algebra can be obtained with only one parameter. Then we use data for the $e^+e^-\to D\bar D$ reaction, from where we determine a form factor needed in the theoretical frame work, as well as other parameters needed to evaluate the meson-meson selfenergy of the $\psi(3770)$. Once this is done we determine the $Z$ probability to still have a vector core and the probability to have the different meson components. We find $Z$ about $80\sim85\%$, and the individual meson-meson components are rather small, providing new empirical information to support the largely $q\bar q$ component of vector mesons, and the $\psi(3770)$ in particular.
\end{abstract}

\maketitle


\section{Introduction}
The nature of hadronic resonances is a field of continuous debate \cite{Crede:2013sze,Chen:2016qju,Chen:2016spr,Guo:2017jvc}. The simple picture of mesons as $q\bar q$ objects and baryons as $qqq$ objects gave an impressive boost to hadron physics and large amount of mesons and baryons were described with this picture \cite{Godfrey:1985xj}. Yet, the advent of a new wave of experiments in the charm and bottom sectors has brought new information that clearly challenges this early picture in many cases \cite{Chen:2016qju,Chen:2016spr,Guo:2017jvc}. Even in the light quark sector there are mesonic resonances that clearly cannot be represented as $q\bar q$ states, as the low lying scalar mesons ($f_0(500)$, $f_0(980)$, $a_0(980)$, $\cdots$) \cite{Oller:1997ti,Kaiser:1998fi,Locher:1997gr,Nieves:1999bx}. On the other hand, the elaborate analysis of meson-meson data by means of QCD and large $N_c$ argument concluded that low lying vector mesons are largely $q\bar q$ objects \cite{Pelaez:2015qba}.

It is unclear whether in the charm or bottom sector one can come to a similar conclusion. In fact, in Ref.~\cite{Barnes:2007xu}  as study was made within the quark model of the meson-meson components of the charmonium vector states, and it was concluded that even the ground state $J/\psi$ had only as survival probability as a vector of about $0.69$ when the meson-meson components to which it couples were considered. This makes us think that higher excited vector charmonium states could actually have even smaller $q\bar q$ components.

In the present work we retake this issue for the $\psi(3770)$ vector state using data from the $e^+e^-\to D\bar D$ reactions. We make an elaborate study of the $D\bar D$, $D\bar D^\ast$, $D^\ast\bar D$, $D^\ast\bar D^\ast$ components of this resonance using the $^3P_0$ model for hadronization of $q\bar q$ into meson-meson components  which requires only one parameter. By means of this and the data of the $e^+e^-\to D^+D^-, D^0\bar D^0$ reactions we can determine the parameters of the theory that allows us to evaluate the meson-meson selfenergy of the $\psi(3770)$. The data of the $e^+e^-\to D\bar D$ reaction are essential for the reliable calculations of the selfenergy, since the unknown couplings and a form factor entering the calculation are extracted from the data. In fact the form factor is relevant to the evaluation of the meson-meson probabilities and we show that it is tied to the fast fall down of the $e^+e^-\to D\bar D$ cross section above the $\psi(3770)$ peak.

The asymmetry of the $\psi(3770)$ peak observed in the $e^+e^-\to D\bar D$ reactions \cite{Ablikim:2008zz,Ablikim:2006zq,Aubert:2006mi} has been the subject of the intense discussion (see Ref.~\cite{Coito:2017ppc} for a recent review). In Ref.~\cite{Coito:2017ppc} a work similar to the one we do here, but using only the $D\bar D$ components, which are the most relevant, is done, and the shape of the $\psi(3770)$ peak is tied to a form factor that is introduced in an empirical way. We also implement this form factor in the same form and two different forms to estimate uncertainties. What we find is that the $\psi(3770)$ is largely a $q\bar q$ state and the meson-meson components are small. The $Z$ probability of having a $q\bar q$ vector core for the $\psi(3770)$ is about $80\sim85\%$ and the individual meson-meson components are small.

This paper is organized as follows. In Sec.~\ref{formalism}, we establish the formalism of calculating the cross section for $e^+e^-\to D\bar D$ through the dressed propagator of $\psi(3770)$, and the meson-meson probabilities in the $\psi(3770)$ wave function. In Sec.~\ref{result}, we present the results on the line shape of $\psi(3770)$ fitting to the experimental data, and then calculate the $Z$ probabilities using the parameters extracted from the fitting. A summary is presented in Sec.~\ref{conclusion}. The angular momentum algebra employed in the calculations is done explicitly in Appendix~\ref{appa}.

\section{Formalism}\label{formalism}
\begin{figure}[h!]
  \centering
  \includegraphics[width=0.60\textwidth]{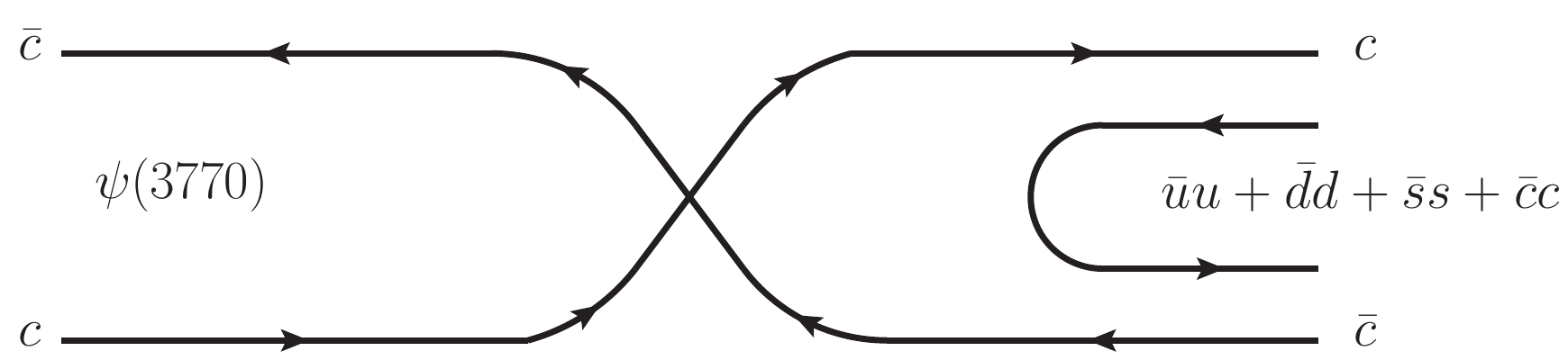}
  \caption{Hadronization process for $\psi(3770)\to D^{(\ast)}\bar D^{(\ast)}$.}
  \label{feyn1}
\end{figure}

Our starting point is the hadronization in the process $\psi\to D^{(\ast)}\bar D^{(\ast)}$ shown in Fig.~\ref{feyn1}, where we introduce a $\bar qq$ pair with the quantum numbers of the vacuum, and insert it between the quark constituents of $\psi(3770)$, $c\bar c$. The insertion of $\bar qq$ is implemented in a $^{3}P_0$ state \cite{Micu:1968mk,LeYaouanc:1972vsx}, which indicates that the inserted $\bar qq$ has positive parity and zero angular momentum, and since $\bar q$ has negative parity we need an orbital angular momentum $L=1$ for $\bar qq$ to fix the parity, which makes $\bar qq$ couple to spin $S=1$, then $S=1$ and $L=1$ couple to total angular momentum $J=0$. The $\psi(3770)$ according to Ref.~\cite{Godfrey:1985xj} corresponds to a $D$-wave $c\bar c$ state with no radial excitation, a $1 ^3D_1$ state with $J^{PC}=1^{--}$.

The hadronization in Fig.~\ref{feyn1} proceeds as follows:
\begin{equation}\label{hadron1}
\psi\to c\bar c\to c(\bar uu+\bar dd+\bar ss+\bar cc)\bar c\to F,
\end{equation}
with $F$
\begin{equation}\label{hadron2}
F=\sum_{i=1}^4 c\bar q_iq_i\bar c=\sum_{i=1}^4M_{4,i}M_{i,4}=(M^2)_{4,4},
\end{equation}
where $M$ corresponds to the following matrix
\begin{equation}\label{matrix1}
M = (q\bar q)=\begin{pmatrix}
u\bar u & u\bar d & u\bar s & u\bar c \\
d\bar u & d\bar d & d\bar s & d\bar c \\
s\bar u & s\bar d & s\bar s & s\bar c \\
c\bar u & c\bar d & c\bar s & c\bar c \\
\end{pmatrix}.
\end{equation}
Alternatively, we can write $q\bar q$ in Eq.~\eqref{matrix1} in terms of their meson components by means of the $\phi$ matrix for pseudoscalar mesons with the mixing between $\eta$ and $\eta^\prime$ taken into account \cite{Bramon:1992kr},
\begin{equation}\label{matrix2}
\phi = \begin{pmatrix}
\frac{1}{\sqrt{2}}\pi^0 + \frac{1}{\sqrt{3}} \eta + \frac{1}{\sqrt{6}}\eta' & \pi^+ & K^+ & \bar{D}^0 \\
 \pi^- & -\frac{1}{\sqrt{2}}\pi^0 + \frac{1}{\sqrt{3}} \eta + \frac{1}{\sqrt{6}}\eta' & K^0 & D^- \\
 K^- & \bar{K}^0 & -\frac{1}{\sqrt{3}} \eta + \sqrt{\frac{2}{3}}\eta' & D_s^- \\
D^0  & D^+ & D_s^+ & \eta_c
\end{pmatrix}.
\end{equation}
Similarly, the vector matrix corresponding to $q\bar q$, which is also needed in our calculations, is given by
\begin{equation}\label{matrix3}
V = \begin{pmatrix}
 \frac{1}{\sqrt{2}}\rho^0 + \frac{1}{\sqrt{2}} \omega & \rho^+ & K^{* +} & \bar{D}^{* 0} \\
 \rho^- & -\frac{1}{\sqrt{2}}\rho^0 + \frac{1}{\sqrt{2}} \omega  & K^{* 0} & \bar{D}^{* -} \\
 K^{* -} & \bar{K}^{* 0}  & \phi & D_s^{* -} \\
 D^{* 0} & D^{* +} & D_s^{* +} & J/\psi
\end{pmatrix}.
\end{equation}

As shown in Eq.~\eqref{hadron2}, where the matrix $M$ could either be the pseudoscalar matrix (which is labeled as $P$ in the following) or the vector matrix (labeled as $V$), we can have four different types of hadronization of the $\psi(3770)$ leading to $PP$, $PV$, $VP$ and $VV$. For example, when both $M$ in Eq.~\eqref{hadron2} are pseudoscalar matrices we have
\begin{equation}\label{phiphi}
(M^2)_{4,4}\to(\phi\phi)_{4,4}=D^0\bar D^0+D^+D^-+D_s^+D_s^-,
\end{equation}
where we have neglected $\eta_c^2$ which is too heavy to be operative in the meson-meson loop that we shall consider below. It can be noticed that, since the $\psi(3770)$ has isospin zero, the final hadronized combination of $D^0\bar D^0+D^+D^-+D^+_sD^-_s$ has isospin zero. Indeed, recalling the isospin doublets
\begin{equation}\label{isospin}
\begin{pmatrix}
D^+\\-D^0
\end{pmatrix},\quad\quad
\begin{pmatrix}
\bar D^0\\D^-
\end{pmatrix},\quad\quad D_s^+,\quad\quad D_s^-,
\end{equation}
Eq.~\eqref{phiphi} can be rewritten in a isospin-zero combination, which is
\begin{equation}
(PP)_{4,4}|I=0\rangle=\sqrt 2|D\bar D,I=0\rangle+|D_s^+D_s^-\rangle.
\end{equation}
\begin{figure}[h!]
  \centering
  \includegraphics[width=0.60\textwidth]{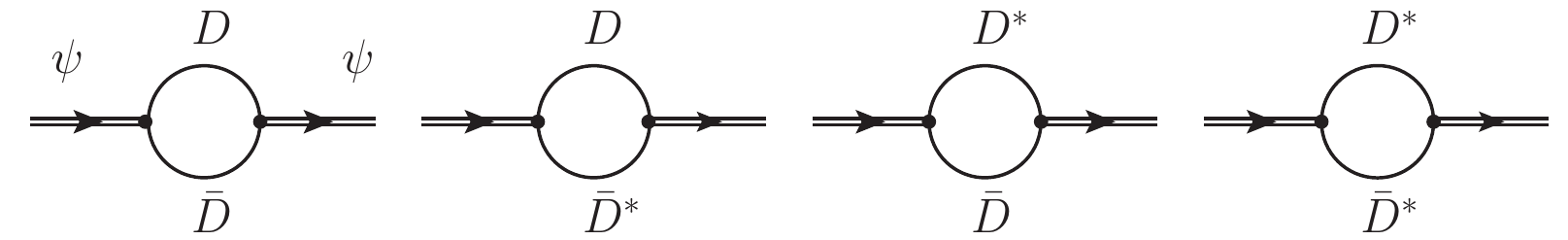}
  \caption{Contribution to the $\psi$ selfenergy for the vector $\psi$ propagator dressed with a meson-meson loop.}
  \label{feyn2}
\end{figure}
Similarly, we can write the combinations coming from $VP$, $PV$ and $VV$
\begin{align}
(PV)_{4,4}&=D^0\bar D^{\ast 0}+D^+D^{\ast-}+D_s^+D_s^{\ast-},\\
(VP)_{4,4}&=D^{\ast 0}\bar D^0+D^{\ast+}D^-+D_s^{\ast+}D_s^-,\\
(VV)_{4,4}&=D^{\ast0}\bar D^{\ast 0}+D^{\ast+}D^{\ast-}+D_s^{\ast+}D_s^{\ast-}.
\end{align}
Note that the combination $(PV)_{4,4}+(VP)_{4,4}$ that we get has the desired negative $C$-parity as it corresponds to the $\psi(3770)$ ($C\,D^\ast=-\bar D^\ast$ in our formalism).

In order to interpret the line shape of the $\psi(3770)$ we follow the steps of Ref.~\cite{Coito:2017ppc}. We consider the propagator of the vector meson $R\equiv\psi(3770)$
\begin{equation}\label{propa1}
G_{\mu\nu}(p)=\left(-g_{\mu\nu}+\frac{p_{\mu}p_{\nu}}{M_R^2}\right)G(p),
\end{equation}
with $G(p)=\frac{1}{p^2-M_R^2+i\varepsilon}$.

The fact that $\psi(3770)$ couples to $PP$, $PV$, $VP$, $VV$ indicates that $\psi(3770)$ will get a selfenergy $\Pi(p)$ that we depict diagrammatically in Fig.~\ref{feyn2}. One can keep the covariant form of $\Pi$, but as shown in Ref.~\cite{Coito:2017ppc} only the transverse part of the propagator is relevant for the discussion here. We argue in a different way, with the same conclusion. In the loop one has $\Pi\sim\int d^4qG(q)G(p-q)$ and the relevant part of it that enters the shape is $Im\Pi$, where the two intermediate mesons are placed on shell. The evaluation of the cross section for $e^+e^-\to D^+D^-$ will place the $D,\bar D$ on shell and the $D$ momenta are about $250\,\rm MeV$. With this small momentum one can neglect the zero component of the $\epsilon^\mu$ polarization vectors. Indeed, as shown in the Appendix of Ref.~\cite{Sakai:2017hpg}, the error induced by neglecting the zero component in this case is $0.7\%$. Hence we need only the spatial component, $\epsilon^i$, and deal with $G_{ij}(p)=\delta_{ij}G(p)\,(i,j=1,2,3)$. When we dress the propagator with the selfenergy of the diagrams in Fig.~\ref{feyn2} we obtain
\begin{align}
G(p)=\frac{1}{p^2-M_R^2-\Pi(p)},
\end{align}
and we must evaluate $\Pi(p)$. Note that we write $M_R$ rather than $M_\psi$ because $M_R$ is now the bare mass of the resonance. The novelty in the present work with respect to Ref.~\cite{Coito:2017ppc} is that we include the contribution of $PV$, $VP$, $VV$ mesons in the selfenergy. They only contribute indirectly to the line shape of the $\psi(3770)$ because $Im\Pi$ is zero in all these cases. However,
\begin{align}
ImG(p)=\frac{Im\Pi(p)}{(p^2-M_R^2-Re\Pi(p))^2+(Im\Pi(p))^2},
\end{align}
and then $Im\Pi$ in the numerator comes only from $D\bar D$, but $Re\Pi(p)$ in the denominator comes from all the channels. Yet, the most novel thing here is that we will evaluate the probability that the $\psi(3770)$ contains $PV$, $VP$ and $VV$ components in its wave function.

The evaluation of $\Pi$ requires to relate the strength of the $PP$, $PV$, $VP$ and $VV$ couplings to the $\psi(3770)$. This we can do with the help of the $^3P_0$ model and the details are given in Appendix~\ref{appa}. While the evaluation is involved, requiring elaborate sums of many Clebsch-Gordan (CG) coefficients, the results are very simple and we write the $\psi(3770)\to PP,PV,VP,VV$ couplings below
\begin{align}
V_{\psi,(MM)_i}=g_{\psi,(MM)_i}\bm\epsilon\,\bm q\,F(\bm q),
\end{align}
with 
\begin{align}\label{coupling1}
g_{\psi,(MM)_i}=A\,C_i\,(i=1,2,3),
\end{align}
and $F(\bm q)$ a form factor coming from the integrals of the quark radial wave functions discussed in Appendix~\ref{appa}, where $A$ in Eq.~\eqref{coupling1} is an unknown coefficient to be fitted to the data, and $C_i$ are the coefficients listed in Table~\ref{tab1}.
\begin{table*}[h!]
\renewcommand\arraystretch{1.0}
\centering
\caption{\vadjust{\vspace{-2pt}}Coefficients $C_i$ for different components in the loop.}\label{tab1}
\begin{tabular*}{\textwidth}{@{\extracolsep{\fill}}c|c|c}
\hline
\hline
 $PP$& $|C_1|^2=\frac{1}{12}$    & $D^+D^-$, $D^0\bar D^0$, $D_s^+D_s^-$ \\
\hline
 $PV,VP$& $|C_2|^2=\frac{1}{6}\times\frac{1}{4}$    & $D^0\bar D^{\ast0}$, $D^{\ast0}\bar D^0$, $D^+\bar D^{\ast-}$, $D^{\ast+}D^-$, $D_s^+D_s^-$, $D_s^+D_s^{\ast-}$, $D_s^{\ast+}D_s^{\ast-}$ \\
\hline
 $VV$& $|C_3|^2=\frac{1}{12}\times\frac{231}{30}$ & $D^{\ast0}\bar D^{\ast0}$, $D^{\ast+}D^{\ast-}$, $D_s^{\ast+}D_s^{\ast-}$\\
\hline\hline
\end{tabular*}
\end{table*}

The former coefficients are for $\psi(3770)$ assumed a $1 ^3D_1$ state.
\begin{figure}[h!]
  \centering
  \includegraphics[width=0.60\textwidth]{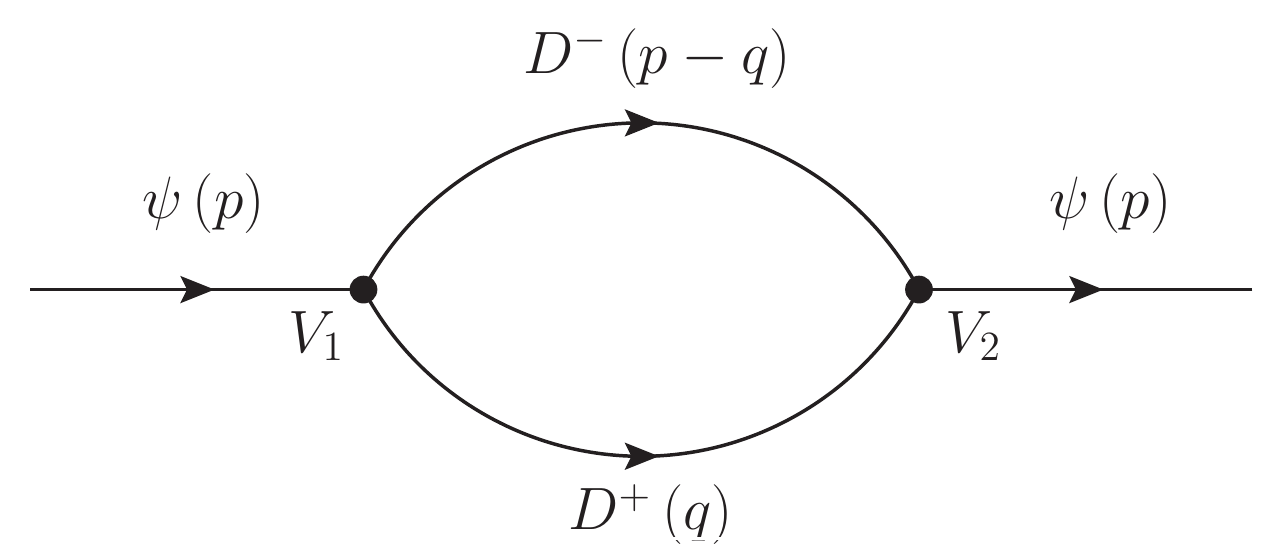}
  \caption{The $\psi$ propagator dressed with a $D^+D^-$ loop as an example.}
  \label{feyn3}
\end{figure}
The terms of the $\Pi(p)$ selfenergy are evaluated as follows, see Fig.~\ref{feyn3}. For $D^+D^-$, for example, we have
\begin{align}
-i\Pi(p)=\int\frac{d^4q}{(2\pi)^4}(-i)V_1(-i)V_2\frac{i}{q^2-m^2_{D^+}+i\varepsilon}\frac{i}{(p-q)^2-m^2_{D^-}+i\varepsilon}F(\bm q)^2,
\end{align}
which gives us
\begin{align}
\Pi(p)=i\,g^2_{\psi,D^+D^-}\int\frac{d^4q}{(2\pi)^4}\bm q^2\frac{1}{q^2-m^2_{D^+}+i\varepsilon}\frac{1}{(p-q)^2-m^2_{D^-}+i\varepsilon}F(\bm q)^2.
\end{align}
The $q^0$ integration can be done analytically and then we get in the rest frame of the $\psi(3770)$ $(p^0=\sqrt s)$
\begin{align}
\Pi(p)=g^2_{\psi,D^+D^-}\tilde G(p^0),
\end{align}
where $\tilde G(p^0)$ has the form
\begin{align}
\tilde G(p^0)&=\int\frac{d^3q}{(2\pi)^3}\frac{1}{2\omega_1(\bm q)}\frac{1}{2\omega_2(\bm q)}\bm q^2\frac{2\omega_1(\bm q)+2\omega_2(\bm q)}{(p^0)^2-(\omega_1(\bm q)+\omega_2(\bm q))^2+i\varepsilon}F(\bm q)^2\nonumber\\
&=\int\frac{dq}{(2\pi)^2}\frac{\omega_1(\bm q)+\omega_2(\bm q)}{\omega_1(\bm q)\omega_2(\bm q)}\frac{\bm q^4}{(p^0)^2-(\omega_1(\bm q)+\omega_2(\bm q))^2+i\varepsilon}F(\bm q)^2,
\end{align}
with $\omega_1(\bm q)=\sqrt{\bm q^2+m^2_{D^+}}$, $\omega_2(\bm q)=\sqrt{\bm q^2+m^2_{D^-}}$.

Let us note in passing that $\tilde G(p^0)$ has a structure similar to the $G(p^0)$ function used in the study of meson-meson interaction \cite{Oller:1997ti} except for the extra factor $\bm q^2$ that makes $\tilde G(p^0)$ more divergent in the absence of the form factor. However, this form factor makes it convergent and we shall come back to it.

With the former expression for $\tilde G(p)$ we can already write the $\psi(3770)$ selfenergy as:
\begin{align}\label{sum}
\Pi(p^0)=&|A|^2\Big\{\frac{1}{12}\tilde G(p^0)\big|_{D^0\bar D^0}+\frac{1}{12}\tilde G(p^0)\big|_{D^+D^-}+\frac{1}{24}\tilde G(p^0)\big|_{D^0\bar D^{\ast0}}+\frac{1}{24}\tilde G(p^0)\big|_{D^{\ast0}\bar D^0}\nonumber\\
&+\frac{1}{24}\tilde G(p^0)\big|_{D^+D^{\ast-}}+\frac{1}{24}\tilde G(p^0)\big|_{D^{\ast+}\bar D^-}+\frac{231}{360}\tilde G(p^0)\big|_{D^{\ast0}\bar D^{\ast0}}+\frac{231}{360}\tilde G(p^0)\big|_{D^{\ast+}D^{\ast-}}\nonumber\\
&+\frac{1}{12}\tilde G(p^0)\big|_{D_s^+D_s^-}+\frac{1}{24}\tilde G(p^0)\big|_{D_s^+D_s^{\ast-}}+\frac{1}{24}\tilde G(p^0)\big|_{D_s^{\ast+}D_s^-}+\frac{231}{360}\tilde G(p^0)\big|_{D_s^{\ast+}D_s^{\ast-}}\Big\}.
\end{align}

Rather than evaluating the form factor $F(q)$ with quark wave function we take an empirical attitude as in Ref.~\cite{Coito:2017ppc}, and let the data determine this form factor from the shape of the $e^+e^-\to D^+D^-$ cross section. Once again we follow Ref.~\cite{Coito:2017ppc} and write
\begin{align}
\sigma=-g^2_{\psi e^+e^-}ImD(M_{inv}),
\end{align}
where $M_{inv}$ is the $e^+e^-$ invariant mass, $\sqrt s$, and $g_{\psi e^+e^-}$, as in Ref.~\cite{Coito:2017ppc}, will also be determined from the strength of the cross section.

It is also useful to separate $\sigma$ into the contribution of the different channels $(D^+D^-,D^0\bar D^0)$. Then we easily write:
\begin{align}
\sigma_i=-g^2_{\psi e^+e^-}ImD_i(M_{inv}),
\end{align}
where
\begin{align}
ImD_i=\frac{Im\Pi_i(p)}{(p^2-M_R^2-Re\Pi(p))^2+(Im\Pi(p))^2},
\end{align}
where $\Pi_i(p)$ is the contribution to $Im\Pi(p^2)$ from the $D^+D^-$ or $D^0\bar D^0$ channel (see Eq.~\eqref{sum}). Note that in the denominator we have $\Pi(p)$, meaning that all channels are included here.

\subsection{Meson-meson probabilities in the $\psi(3770)$ wave function}
Let us write for convenience, as in Ref.~\cite{Coito:2017ppc},
\begin{align}
\Pi^\prime(p)=\Pi(p)-Re(\Pi(M_\psi)),
\end{align}
which vanishes at $\sqrt s=M_\psi$, and with this choice we can write
\begin{align}
G(p)=\frac{1}{p^2-M^2_\psi-\Pi^\prime(p)}.
\end{align}
We can make an expansion around $M_\psi$ and have
\begin{align}
G(p)&=\frac{1}{p^2-M^2_\psi-Re(\Pi^\prime(p))-iIm\Pi(p)}\nonumber\\
&=\frac{1}{p^2-M^2_\psi-[Re(\Pi^\prime(p))-Re(\Pi^\prime(M_\psi))]-iIm\Pi(p)},
\end{align}
since $Re\Pi^\prime(M_\psi)=0$ and hence
\begin{align}\label{g1}
G(p)&\simeq\frac{1}{p^2-M^2_\psi-\frac{\partial Re\Pi}{\partial p^2}\big|_{M^2_\psi}(p^2-M^2_\psi)-iIm\Pi(p)}\nonumber\\
&=\frac{1}{(p^2-M^2_\psi)(1-\frac{\partial Re\Pi}{\partial p^2}\big|_{M^2_\psi})-iIm\Pi(p)}\nonumber\\
&=\frac{Z}{p^2-M^2_\psi-iZIm(p)},
\end{align}
with
\begin{align}\label{proba1}
Z&=\frac{1}{1-\frac{\partial Re\Pi(p^2)}{\partial p^2}\Big|_{p^2=M^2_\psi}}\nonumber\\
&\simeq 1+\frac{\partial Re\Pi}{\partial p^2}\Big|_{p^2=M^2_\psi}.
\end{align}

This is the typical wave function renormalization \cite{Itzykson:1980rh} and $Z$ is interpreted as the probability to still have the original vector when it is dressed by the meson-meson components. Conversely $1-Z$ will be the meson-meson probability of the dressed vector. If $\frac{\partial Re\Pi}{\partial p^2}$ is reasonably smaller than $1$, one can make an expansion as in Eq.~\eqref{proba1}, and furthermore we have
\begin{align}
1-Z=-\frac{\partial Re\Pi}{\partial p^2}\Big|_{p^2=M^2_\psi},
\end{align}
such that $-\frac{\partial Re\Pi}{\partial p^2}\Big|_{p^2=M^2_\psi}$ can be interpreted as the meson-meson probability and in particular one can get the contribution of each channel:
\begin{align}\label{proba2}
P_{(MM)i}\simeq-\frac{\partial Re\Pi_i(p^2)}{\partial p^2}\Big|_{p^2=M^2_\psi},
\end{align}
where $\Pi_i$ is the contribution of $i$-th channel to $\Pi$.

\section{Results}\label{result}
In Ref.~\cite{Coito:2017ppc} a form factor was used
\begin{align}\label{ff0}
f_\Lambda(\xi)=e^{-\xi/(4\Lambda^2)}\,e^{(m^2_{D^0}+m^2_{D^+})/(2\Lambda^2)},
\end{align}
with $\xi=4(\bm q^2+m^2)$, that is the equivalent to our $F(\bm q)^2$, and $\Lambda$ was fitted to data. We get similar results using this form factor. In addition, we use two other form factors:
\begin{align}\label{ff1}
F(\bm q)^2=\frac{1+(R\,q_{on})^2}{1+(R\,q)^2},
\end{align}
and
\begin{align}\label{ff2}
F(\bm q)^2=\frac{1+(R\,q_{on})^4}{1+(R\,q)^4},
\end{align}
with $q_{on}$ the following form for $D\bar D$
\begin{align}
q_{on}=\frac{\lambda^{1/2}(M^2_\psi,m_D^2,m_{\bar D}^2)}{2M_\psi},
\end{align}
where $\lambda$ is the usual K\"{a}ll\'en function, and the parameter $R$ is fitted to the data in both cases. We have thus four parameters, as in Ref.~\cite{Coito:2017ppc}, which in our case are $M_{\psi}$, $g_{\psi e^+e^-}$, $A$ and $R$. $M_{\psi}$ is of course very close to the nominal mass of the $\psi(3770)$, $g_{\psi e^+e^-}$ determines the strength of the cross section, $A$ is related to the width of the resonance, and $R$ determines the fall down of the resonance shape above the resonance peak. The parameters are fitted to the data of the cross section for $e^+e^-\to D\bar D$ \cite{Ablikim:2008zz,Ablikim:2006zq,Aubert:2006mi}.

Given the fact that in the Appendix~\ref{appa} we found that the form factor comes from an integral of the radial wave function of the quarks, and these are the same, independent of the different spin couplings, we assume this form factor to be the same for the $PV$, $VP$ and $VV$ cases.
\begin{figure}[h!]
  \centering
  \includegraphics[width=0.60\textwidth]{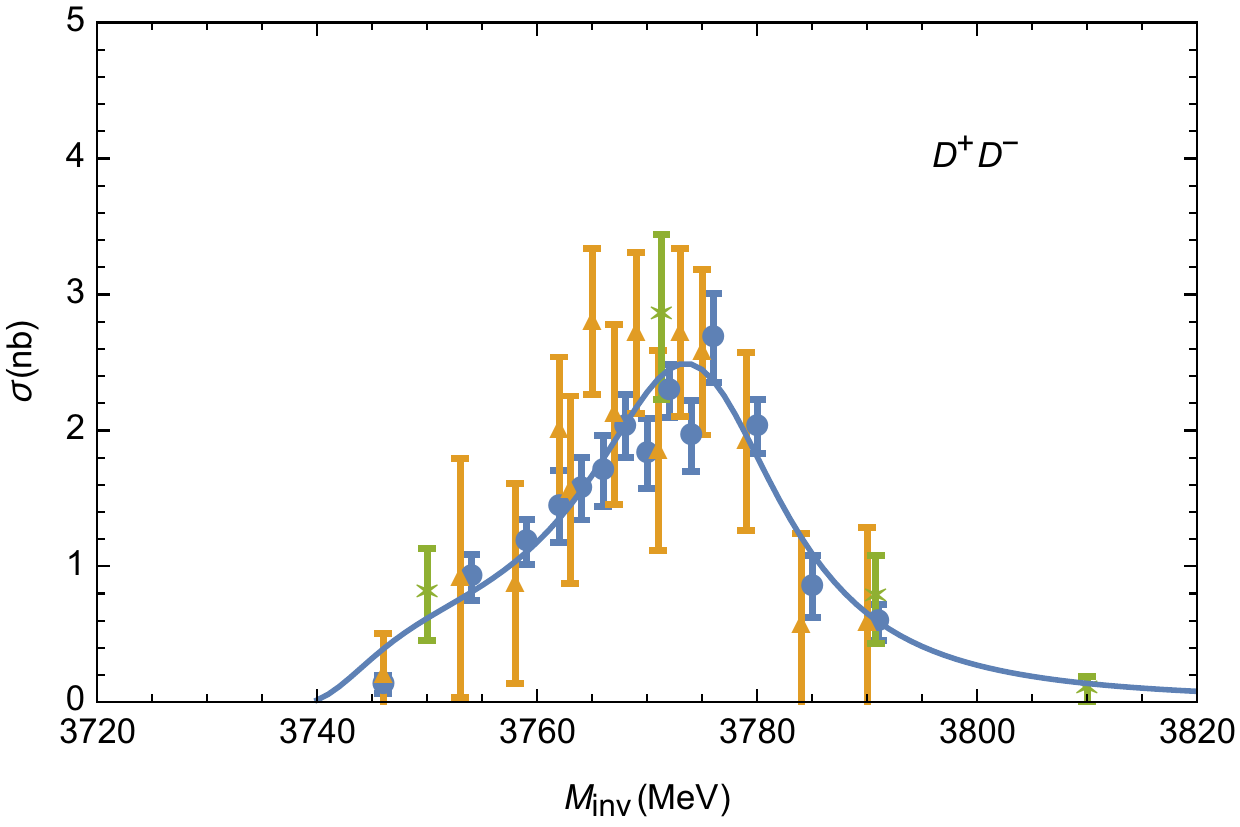}
  \caption{Cross section of $e^+e^-\to D^+D^-$ fitted to the experimental data ($\bullet$\cite{Ablikim:2008zz}, $\blacktriangle$\cite{Ablikim:2006zq}, $\star$\cite{Aubert:2006mi}) using the form factor of Eq.~\eqref{ff2}.}
  \label{fig4}
\end{figure}

In Fig.~\ref{fig4} we show the results for the $e^+e^-\to D^+D^-$ cross section using the form factor of Eq.~\eqref{ff2}. The parameters used can be seen in Table~\ref{tab2}. As we can see, there is a good fit to the data, both above and below the peak, reflecting the asymmetry of the distribution, which does not have a Breit-Wigner form.
\begin{table*}[h!]
\renewcommand\arraystretch{1.3}
\centering
\caption{\vadjust{\vspace{-2pt}}Fitting parameters for Fig.~\ref{fig4}.}\label{tab2}
\begin{tabular*}{0.50\textwidth}{@{\extracolsep{\fill}}cc}
\hline
\hline
 $M_R$    &   $3773\,\rm MeV$  \\
\hline
 $g^2_{\psi e^+e^-}$& $1.40\times10^{-6}$  \\
\hline
 $R$      &$0.0070\,\rm MeV^{-1}$  \\
\hline
 $|A|^2$    &   1750\\
\hline
\hline
\end{tabular*}
\end{table*}

We should note that the description of the data is a result of the parametrization, and in particular the fall down of the distribution above the peak is related to the parameter $R$. There is nothing fundamental in this interpretation of the asymmetry. However, the data and particularly the fall down above the threshold determine the range of the form factor, and this is important to make the integral $\tilde G(p)$ convergent, such that the probabilities that we obtain are a consequence of the peculiar shape of the $e^+e^-\to D^+D^-$ data. In this sense, the probabilities that we obtain are a prediction based on the $e^+e^-\to D^+D^-$ data, while those in Ref.~\cite{Barnes:2007xu} were based on a particular quark model.
\begin{figure}[h!]
  \centering
  \includegraphics[width=0.60\textwidth]{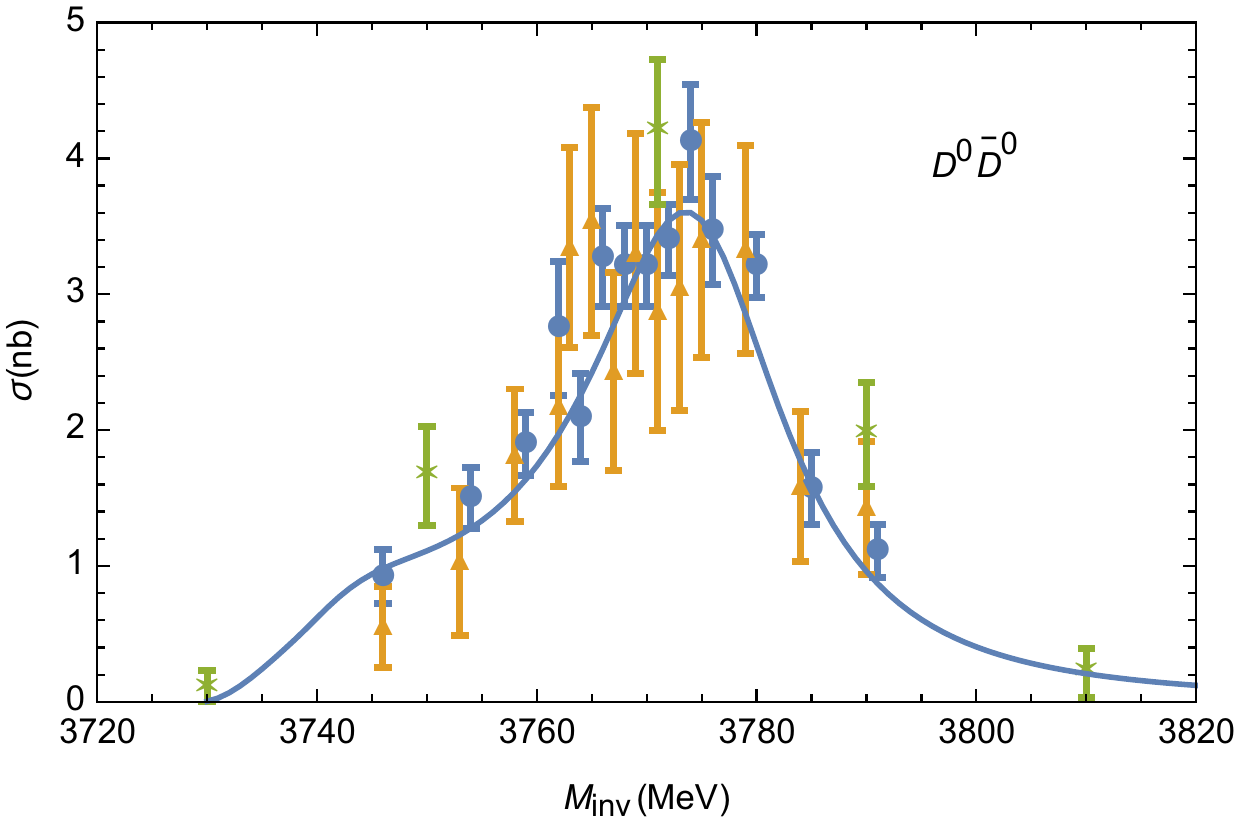}
  \caption{The comparison of our result with the experimental data ($\bullet$\cite{Ablikim:2008zz}, $\blacktriangle$\cite{Ablikim:2006zq}, $\star$\cite{Aubert:2006mi}) for the cross section of $e^+ e^- \to D^0\bar D^0$ reaction, using the form factor of Eq.~\eqref{ff2} and the parameters in Table~\ref{tab2}.}
  \label{fig5}
\end{figure}
\begin{figure}[h!]
  \centering
  \includegraphics[width=0.60\textwidth]{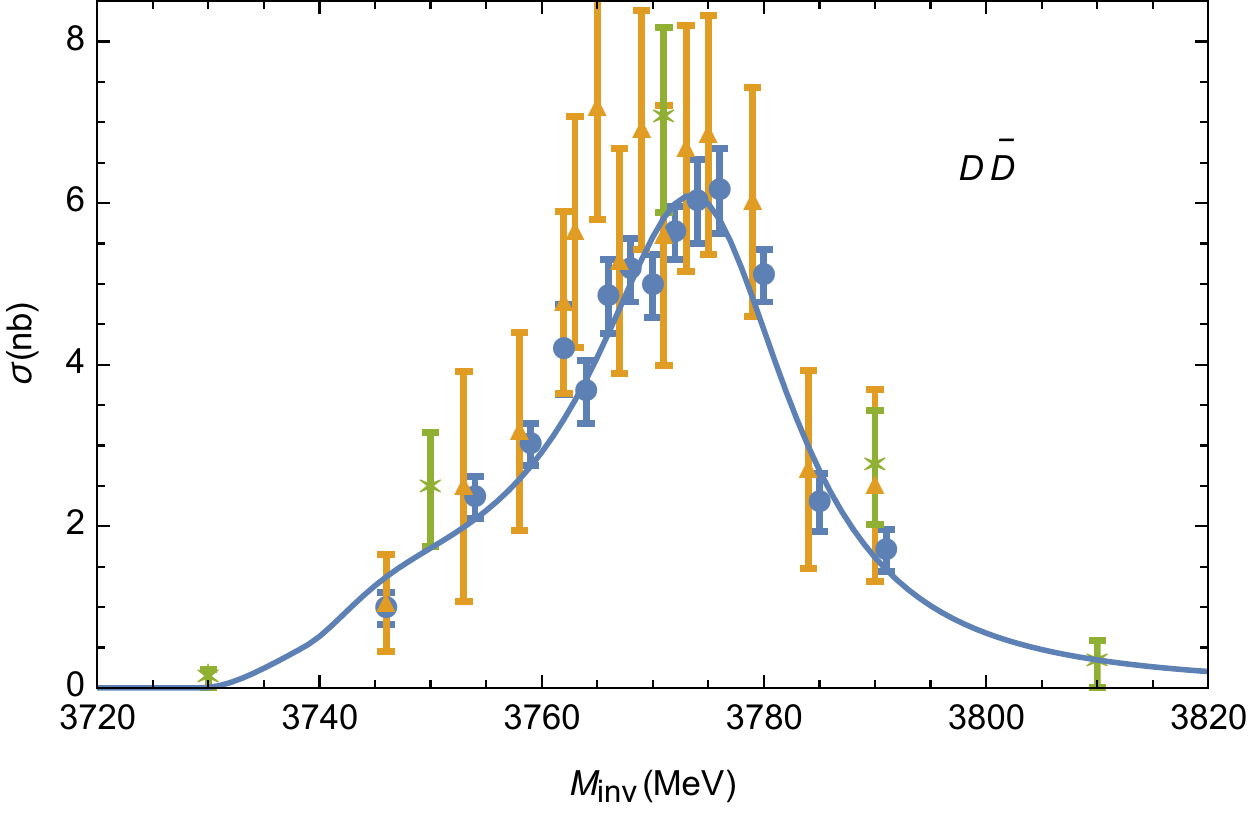}
  \caption{The comparison of our result with the experimental data ($\bullet$\cite{Ablikim:2008zz}, $\blacktriangle$\cite{Ablikim:2006zq}, $\star$\cite{Aubert:2006mi}) for the cross section of $e^+e^-\to D^+D^-+D^0\bar D^0$ reaction, using the form factor of Eq.~\eqref{ff2} and the parameters in Table~\ref{tab2}.}
  \label{fig6}
\end{figure}

\begin{table*}[tbh!]
\renewcommand\arraystretch{1.3}
\centering
\caption{\vadjust{\vspace{-2pt}}Fitting parameters for Fig.~\ref{fig7}.}\label{tab3}
\begin{tabular*}{0.50\textwidth}{@{\extracolsep{\fill}}cc}
\hline
\hline
 $M_R$    &   $3773\,\rm MeV$  \\
\hline
 $g^2_{\psi e^+e^-}$& $1.55\times10^{-6}$  \\
\hline
 $R$      &$0.0030\,\rm MeV^{-1}$  \\
\hline
 $|A|^2$    &   2756\\
\hline
\hline
\end{tabular*}
\end{table*}
It is also interesting to evaluate the $e^+e^-\to D^0\bar D^0$ cross section and compare with the data, This is done in Fig.~\ref{fig5}. We can see that the agreement with the data is also very good, Note that once the $e^+e^-\to D^+D^-$ is fitted, we have no freedom for the $e^+e^-\to D^0\bar D^0$, so the latter one is a prediction of the approach.

In Fig.~\ref{fig6} we show the result for the $e^+e^-\to D^+D^-+D^0\bar D^0$. Obviously, since the individual cross sections are well produced, so is the sum of the two.

\begin{figure}[h!]
  \centering
  \includegraphics[width=0.60\textwidth]{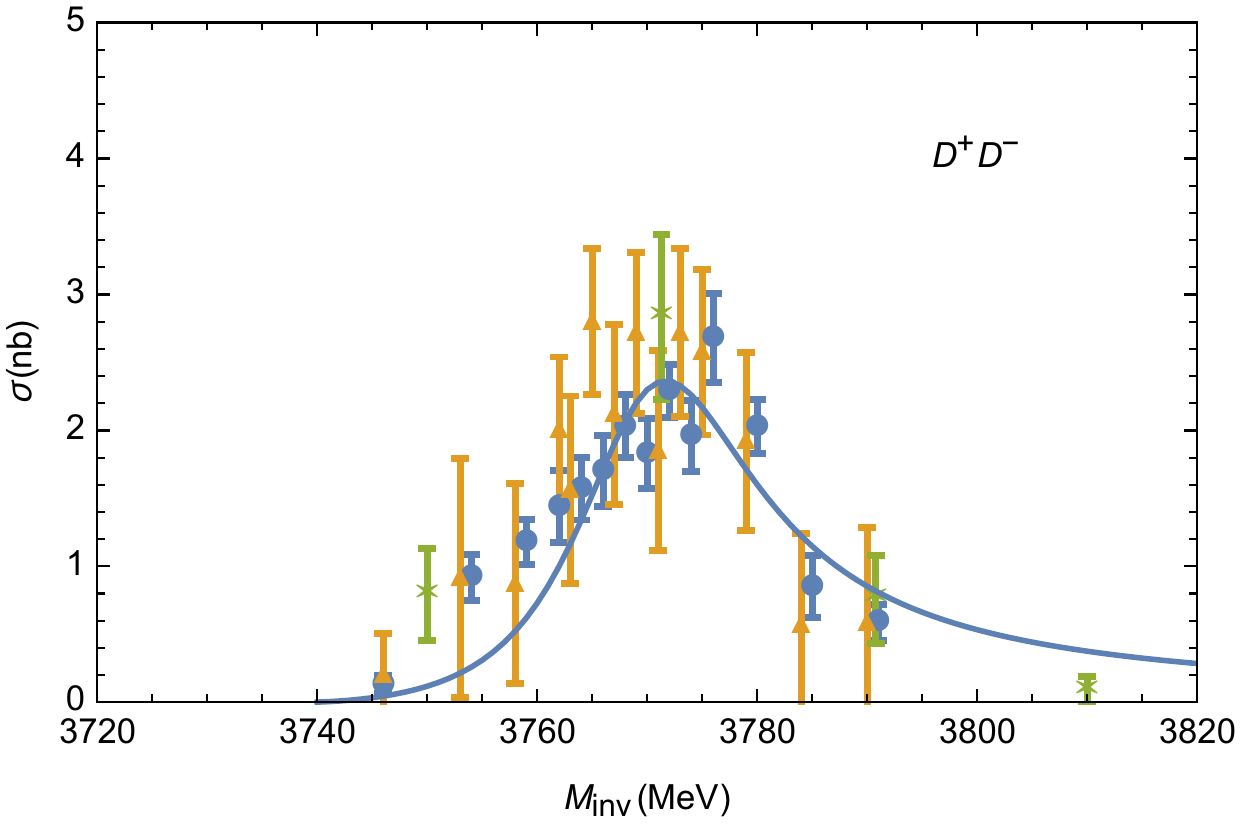}
  \caption{Cross section of $e^+e^-\to D^+D^-$ fitted to the experimental data ($\bullet$\cite{Ablikim:2008zz}, $\blacktriangle$\cite{Ablikim:2006zq}, $\star$\cite{Aubert:2006mi}) using the form factor of Eq.~\eqref{ff1}.}
  \label{fig7}
\end{figure}

\begin{figure}[h!]
  \centering
  \includegraphics[width=0.60\textwidth]{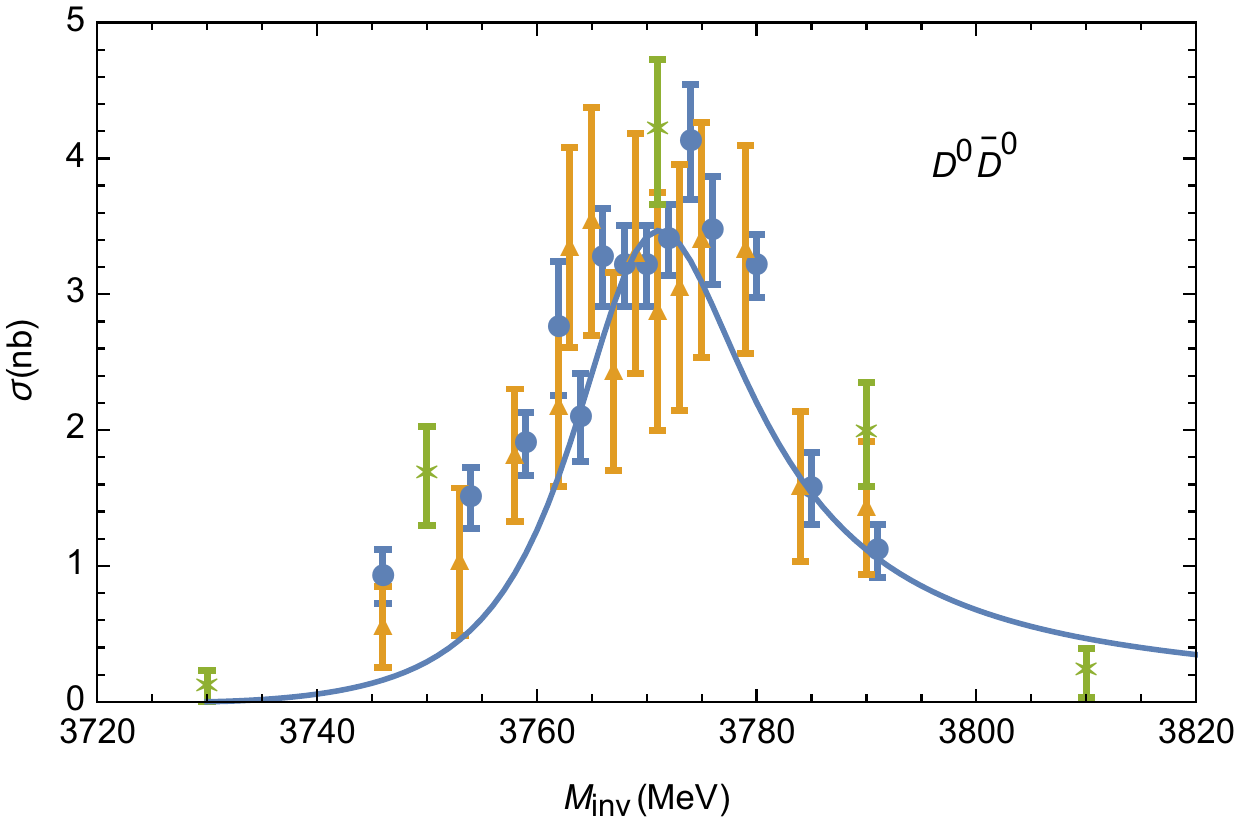}
  \caption{The comparison of our result with the experimental data ($\bullet$\cite{Ablikim:2008zz}, $\blacktriangle$\cite{Ablikim:2006zq}, $\star$\cite{Aubert:2006mi}) for the cross section of $e^+ e^- \to D^0\bar D^0$ reaction, using the form factor of Eq.~\eqref{ff1} and the parameters in Table~\ref{tab3}.}
  \label{fig8}
\end{figure}

\begin{figure}[h!]
  \centering
  \includegraphics[width=0.60\textwidth]{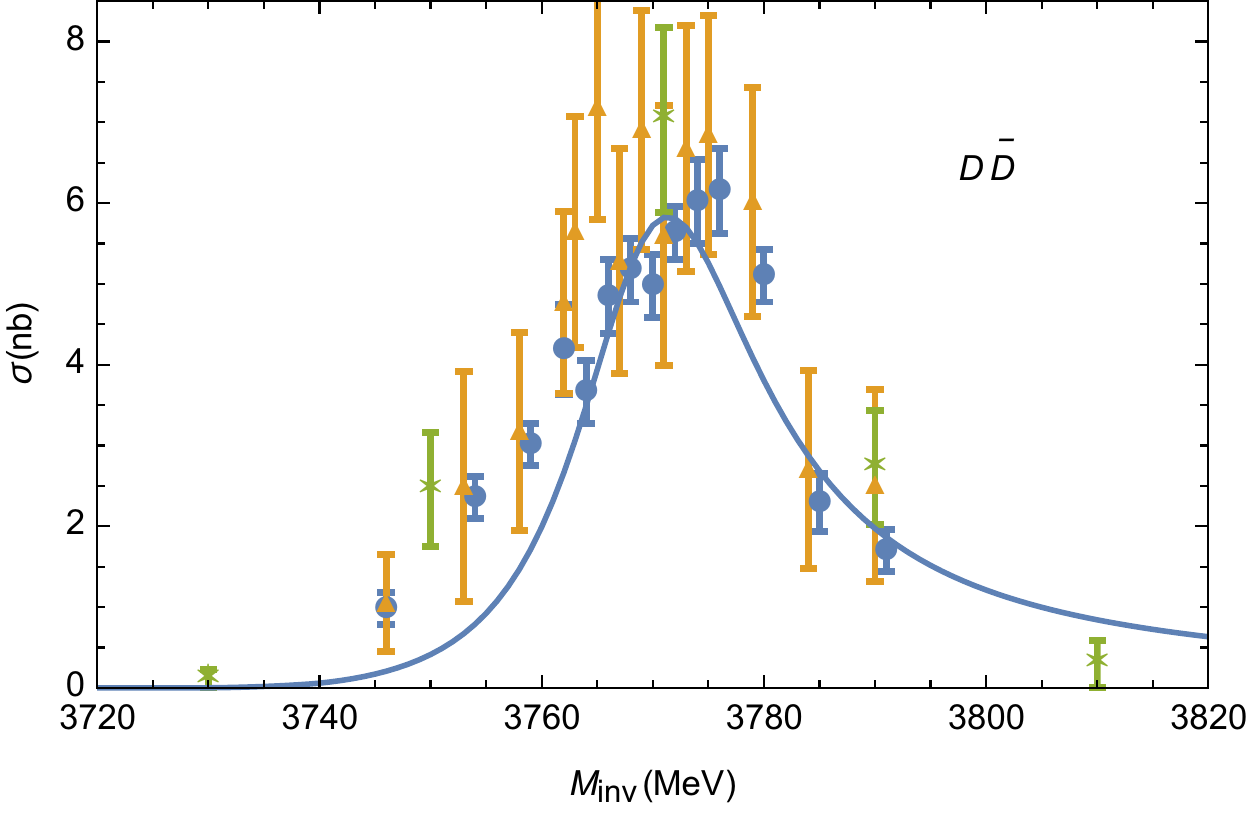}
  \caption{The comparison of our result with the experimental data ($\bullet$\cite{Ablikim:2008zz}, $\blacktriangle$\cite{Ablikim:2006zq}, $\star$\cite{Aubert:2006mi}) for the cross section of $e^+e^-\to D^+D^-+D^0\bar D^0$ reaction, using the form factor of Eq.~\eqref{ff1} and the parameters in Table~\ref{tab3}.}
  \label{fig9}
\end{figure}

Next we show the result of the calculations using the form factor of Eq.~\eqref{ff1}. The parameters of the fit are shown in Table~\ref{tab3}. The result for $e^+e^-\to D^+D^-$, $e^+e^-\to D^0\bar D^0$ and $e^+e^-\to D^+D^-+D^0\bar D^0$ are shown in Figs.~\ref{fig7}, \ref{fig8}, \ref{fig9}. We observe a good fit in the region above the peak, but not as good as before below it, although still comparable with the bulk of the data. Concerning our main goal, which is the evaluation of the meson-meson probabilities, the fall down of the cross section above the peak is acceptable.

\subsection{Evaluation of the vector and meson-meson probabilities}
In Table~\ref{tab4} we show the probability of Eqs.~\eqref{proba1} and \eqref{proba2} using the form factor of Eq.~\eqref{ff2}.
\begin{table*}[h!]
\renewcommand\arraystretch{1.2}
\centering
\caption{\vadjust{\vspace{-2pt}}Meson-meson probabilities in the $\psi(3770)$ wave function with the form factor of Eq.~\eqref{ff2}.}\label{tab4}
\begin{tabular*}{0.70\textwidth}{@{\extracolsep{\fill}}cccc}
\hline
\hline
   Channels                  & $-\frac{\partial\Pi}{\partial p^2}\big|_{p^2=M^2_\psi}$ &$P_{(MM)}$                   & $Z$\\
                               \hline
    $D^0\bar D^0$            &$-0.0555-0.0406i$  &$-0.0555$                    &1.059\\
    \hline
    $D^+D^-$                 &$-0.0879-0.0444i$  &$-0.0879$                    &1.096\\
    \hline
    $D^0\bar D^{\ast0}+c.c$  &$0.0083$  &$0.0083$                     &0.992\\
    \hline
    $D^+\bar D^{\ast-}+c.c$  &$0.0074$  &$0.0074$                     &0.993\\
    \hline
    $D^{\ast0}\bar D^{\ast0}$&$0.0164$  &$0.0164$                     &0.984\\
    \hline
    $D^{\ast+}D^{\ast-}$     &$0.0156$  &$0.0156$                     &0.985\\
    \hline
    $D^+_sD^-_s$             &$0.0040$  &$0.0040$                     &0.996\\
    \hline
    $D_s^+D_s^{\ast-}+c.c$   &$0.0014$  &$0.0014$                     &0.999\\
    \hline
    $D^{\ast+}_sD^{\ast-}_s$ &$0.0054$  &$0.0054$                     &0.995\\
    \hline
    Total                    &$-0.0850-0.0846i$  &$-0.0850$                    &1.093\\
\hline
\hline
\end{tabular*}
\end{table*}
What we see is that the probabilities of the $D^+D^{\ast-}+c.c$ or $D^0D^{\ast0}+c.c$ are practically zero. However, there is the unpleasant feature that $-\frac{\partial\Pi_{D\bar D}}{\partial p^2}\big|_{p^2=M^2_\psi}$ is complex, and $-\frac{\partial Re\Pi_{D\bar D}}{\partial p^2}\big|_{p^2=M^2_\psi}$ $(P_{(MM)})$ is negative. The complex value is unavoidable when one has open channels, but that $-\frac{\partial Re\Pi_{D\bar D}}{\partial p^2}\big|_{p^2=M^2_\psi}$, which provides the $D\bar D$ probability as we have seen, is negative, is unexpected and unacceptable. Fortunately, the value is very small, and could be admitted as an uncertainty related to the approximation implicit in Eq.~\eqref{g1}. As a consequence of this negative number, the $Z$ probability of having the original vector in the $\psi(3770)$ wave function is bigger than one. Yet, by an amount of $9.3\%$, which tells us the uncertainties that we have in this approach. It is interesting to note that if we use the form factor of Ref.~\cite{Coito:2017ppc}  written in Eq.~\eqref{ff0} we get similar results.

In view of this, we use a form factor more in agreement with phenomenology, which is the one of Eq.~\eqref{ff1}. This form factor induces a correction to the width
\begin{align}
\Gamma(s)\to\Gamma_0\frac{1+(R\,q_{on})^2}{1+(R\,\bar q)^2},
\end{align}
with
\begin{align}
\bar q=\frac{\lambda^{1/2}(s,m_D^2,m_{\bar D}^2)}{2\sqrt s},
\end{align}
where $\Gamma_0$ is the width evaluated at $\sqrt s=M_\psi$. This factor is the Blatt-Weisskopf barrier penetration factor \cite{Blatt:1952ije}, commonly used to write the width in usual Breit-Wigner amplitudes. In view of this, we can give more credit to the results that come from this factor. The results can be seen in Table~\ref{tab5}.
\begin{table*}[h!]
\renewcommand\arraystretch{1.2}
\centering
\caption{\vadjust{\vspace{-2pt}}Meson-meson probabilities in the $\psi(3770)$ wave function with the form factor of Eq.~\eqref{ff1} (Note that the sum of the total $P_{(MM)}$ and $Z$ is not exactly $1$ because of the approximation of Eq.~\eqref{proba1}).}\label{tab5}
\begin{tabular*}{0.70\textwidth}{@{\extracolsep{\fill}}cccc}
\hline
\hline
   Channels                  &$-\frac{\partial\Pi}{\partial p^2}\big|_{p^2=M^2_\psi}$ &$P_{(MM)}$                   & $Z$\\
                               \hline
    $D^0\bar D^0$            &$0.0019+0.1814i$       &$0.0019$       &0.998\\
    \hline
    $D^+D^-$                 &$0.0295+0.1862i$       &$0.0295$       &0.971\\
    \hline
    $D^0\bar D^{\ast0}+c.c$  &$0.0264+0.0003i$       &$0.0264$       &0.974\\
    \hline
    $D^+\bar D^{\ast-}+c.c$  &$0.0244+0.0002i$       &$0.0244$       &0.976\\
    \hline
    $D^{\ast0}\bar D^{\ast0}$&$0.0708+0.0004i$       &$0.0708$       &0.934\\
    \hline
    $D^{\ast+}D^{\ast-}$     &$0.0681+0.0004i$       &$0.0681$       &0.936\\
    \hline
    $D^+_sD^-_s$             &$0.0152+0.0001i$       &$0.0152$       &0.985\\
    \hline
    $D_s^+D_s^{\ast-}+c.c$   &$0.0065$               &$0.0065$       &0.994\\
    \hline
    $D^{\ast+}_sD^{\ast-}_s$ &$0.0268$               &$0.0268$       &0.974\\
    \hline
    Total                    &$0.2696+0.3690i$       &$0.2696$       &0.787\\
\hline
\hline
\end{tabular*}
\end{table*}

Now we can see that all the probabilities are positive and the $Z$ probability is smaller than one. Yet, the results that one obtains indicate small meson-meson probabilities and a total probability for $Z$ to have still a vector component is about $80\%$.

\begin{figure}[h!]
  \centering
  \includegraphics[width=0.60\textwidth]{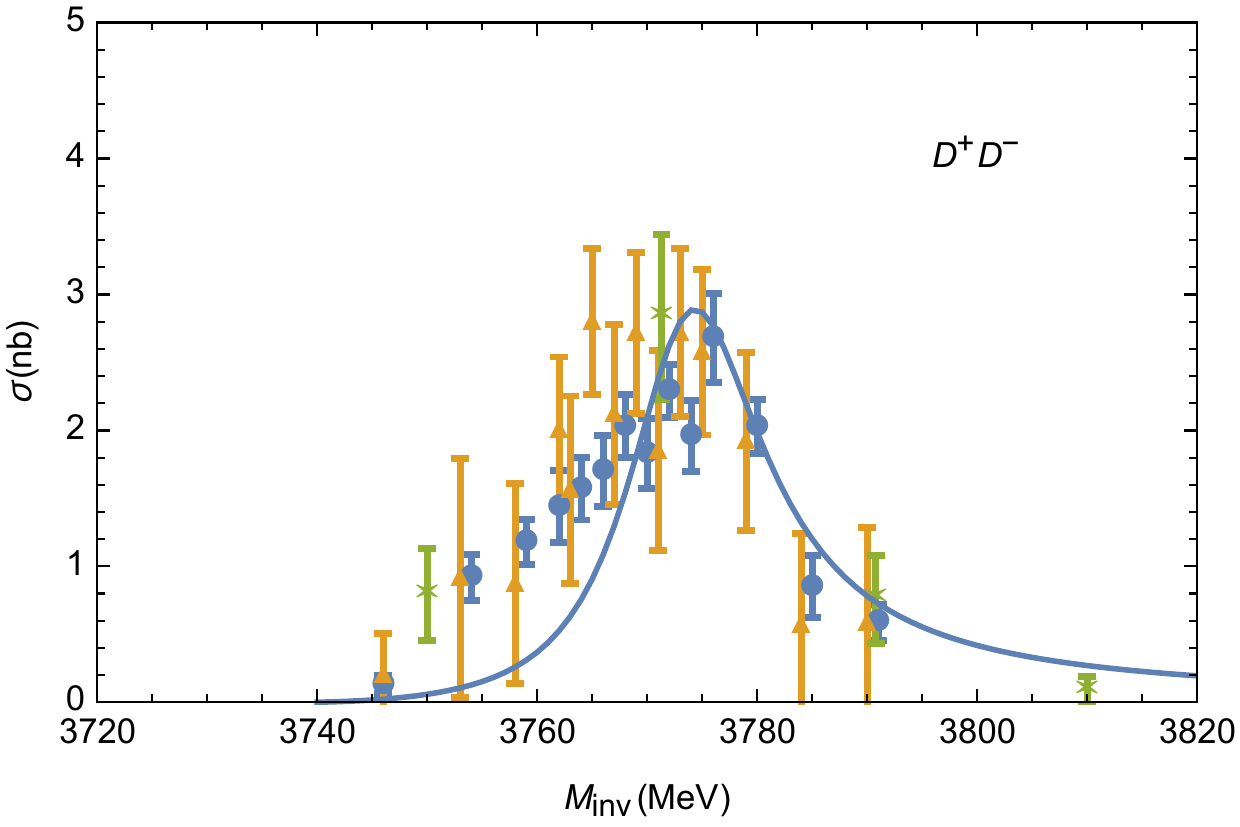}
  \caption{Cross section of $e^+e^-\to D^+D^-$ fitted to the experimental data ($\bullet$\cite{Ablikim:2008zz}, $\blacktriangle$\cite{Ablikim:2006zq}, $\star$\cite{Aubert:2006mi}) using the form factor of Eq.~\eqref{ff1}.}
  \label{fig10}
\end{figure}

\begin{figure}[h!]
  \centering
  \includegraphics[width=0.60\textwidth]{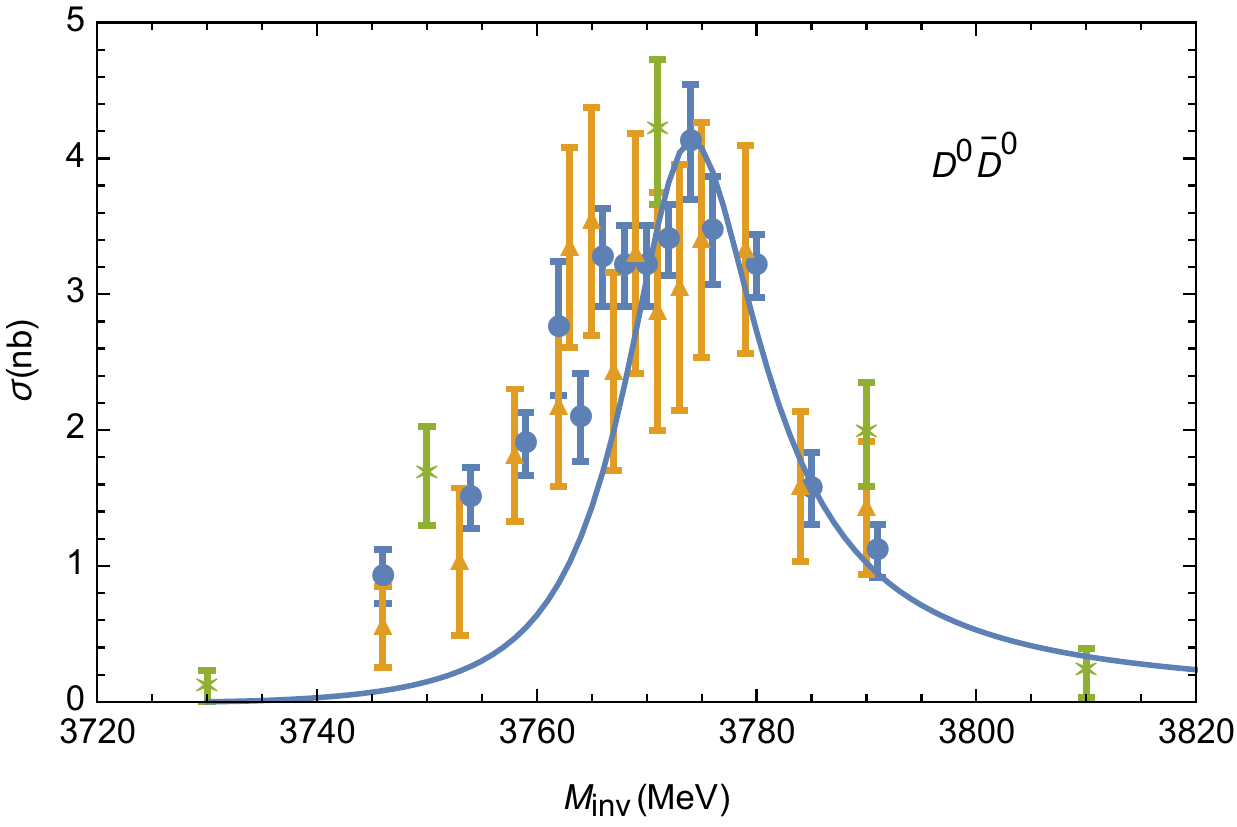}
  \caption{The comparison of our result with the experimental data ($\bullet$\cite{Ablikim:2008zz}, $\blacktriangle$\cite{Ablikim:2006zq}, $\star$\cite{Aubert:2006mi}) for the cross section of $e^+ e^- \to D^0\bar D^0$ reaction, using the form factor of Eq.~\eqref{ff1} and the parameters in Table~\ref{tab6}.}
  \label{fig11}
\end{figure}

\begin{figure}[h!]
  \centering
  \includegraphics[width=0.60\textwidth]{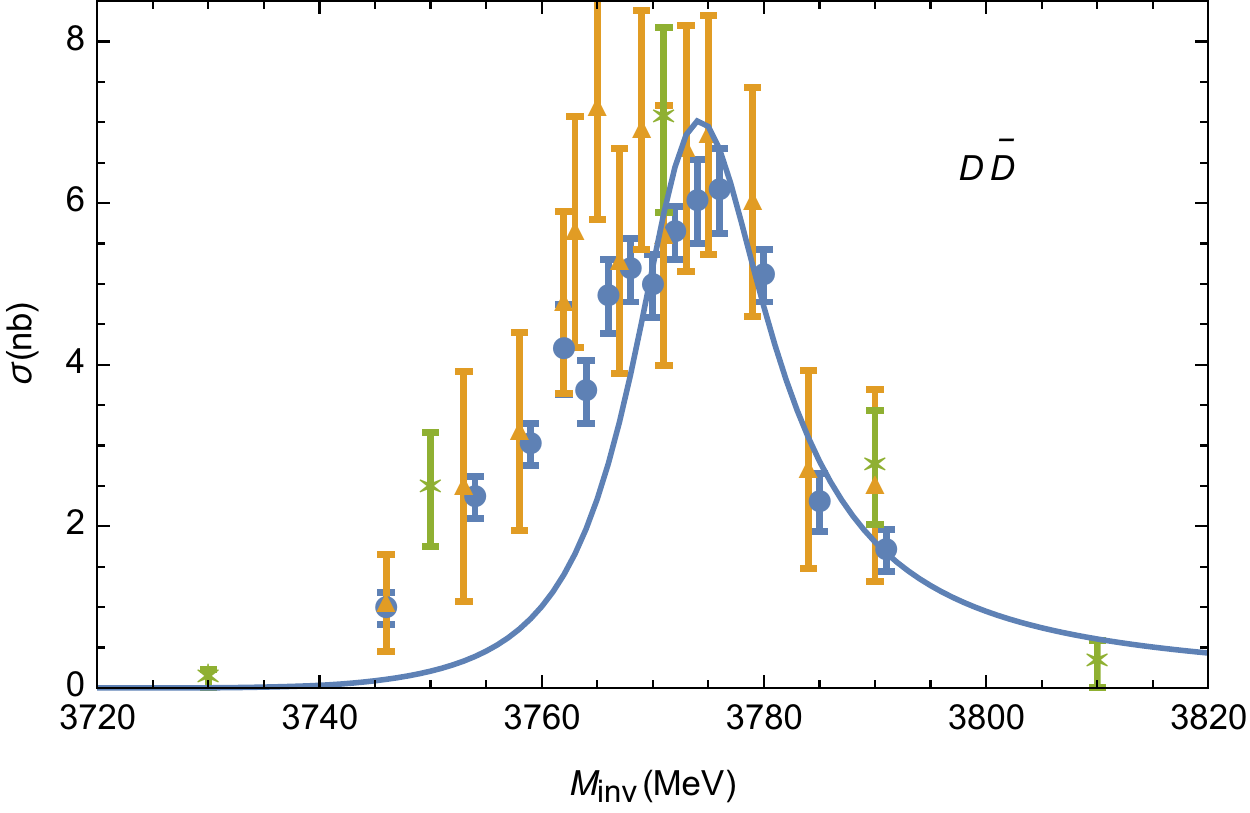}
  \caption{The comparison of our result with the experimental data ($\bullet$\cite{Ablikim:2008zz}, $\blacktriangle$\cite{Ablikim:2006zq}, $\star$\cite{Aubert:2006mi}) for the cross section of $e^+e^-\to D^+D^-+D^0\bar D^0$ reaction, using the form factor of Eq.~\eqref{ff1} and the parameters in Table~\ref{tab6}.}
  \label{fig12}
\end{figure}

We can see that in Figs.~\ref{fig7}, \ref{fig8}, evaluated with the form factor of Eq.~\eqref{ff1} the slope of the cross section above the peak is smaller than in the corresponding Figs.~\ref{fig4}, \ref{fig5}, evaluated with the form factor of Eq.~\eqref{ff2}. We stated our preference for the form factor of Eq.~\eqref{ff1}, more in agreement with phenomenology. In view of that we choose a different set of parameters that make the slope above the peak more similar in all cases, paying the price of not having such good agreement at low energies. However, for the meson-meson probabilities that we are concerned about, the slope above the peak is what matters. The parameters of such a set are shown in Table~\ref{tab6},
\begin{table*}[h!]
\renewcommand\arraystretch{1.3}
\centering
\caption{\vadjust{\vspace{-2pt}}Fitting parameters for Fig.~\ref{fig10}.}\label{tab6}
\begin{tabular*}{0.50\textwidth}{@{\extracolsep{\fill}}cc}
\hline
\hline
 $M_R$    &   $3775\,\rm MeV$  \\
\hline
 $g^2_{\psi e^+e^-}$& $1.25\times10^{-6}$  \\
\hline
 $R$      &$0.0029\,\rm MeV^{-1}$  \\
\hline
 $|A|^2$    &   1700\\
\hline
\hline
\end{tabular*}
\end{table*}
and the results are shown in Figs.~\ref{fig10}, \ref{fig11}, \ref{fig12} and Table~\ref{tab7}. In this case we find $Z\sim0.854$. This is a reasonable number, but in view of the results in Table~\ref{tab5} with the former fit, we can settle the value of $Z$ within $0.80\sim0.85$, which is a reasonable range of uncertainty.
\begin{table*}[h!]
\renewcommand\arraystretch{1.2}
\centering
\caption{\vadjust{\vspace{-2pt}}Meson-meson probabilities in the $\psi(3770)$ wave function with the form factor of Eq.~\eqref{ff1}.}\label{tab7}
\begin{tabular*}{0.70\textwidth}{@{\extracolsep{\fill}}cccc}
\hline
\hline
   Channels                  &$-\frac{\partial\Pi}{\partial p^2}\big|_{p^2=M^2_\psi}$ &$P_{(MM)}$                   & $Z$\\
                               \hline
    $D^0\bar D^0$            &$0.0001+0.1150i$       &$0.0019$       &0.998\\
    \hline
    $D^+D^-$                 &$0.0168+0.1178i$       &$0.0295$       &0.971\\
    \hline
    $D^0\bar D^{\ast0}+c.c$  &$0.0172+0.0002i$       &$0.0264$       &0.974\\
    \hline
    $D^+\bar D^{\ast-}+c.c$  &$0.0158+0.0001i$       &$0.0244$       &0.976\\
    \hline
    $D^{\ast0}\bar D^{\ast0}$&$0.0458+0.0003i$       &$0.0708$       &0.934\\
    \hline
    $D^{\ast+}D^{\ast-}$     &$0.0440+0.0002i$       &$0.0681$       &0.936\\
    \hline
    $D^+_sD^-_s$             &$0.0098$               &$0.0152$       &0.985\\
    \hline
    $D_s^+D_s^{\ast-}+c.c$   &$0.0042$               &$0.0065$       &0.994\\
    \hline
    $D^{\ast+}_sD^{\ast-}_s$ &$0.0172$               &$0.0268$       &0.974\\
    \hline
    Total                    &$0.1709+0.2336i$       &$0.1709$       &0.854\\
\hline
\hline
\end{tabular*}
\end{table*}

This result is very valuable and we consider it the most important output of the work. There is a continuous debate about the nature of the hadron resonances and it is long since the ideal picture of mesons as pure $q\bar q$ and baryons as $qqq$ has been abandoned. With the advent of hadrons in the charm and bottom sectors, the evidence for more complex structures is appalling \cite{Chen:2016qju,Chen:2016spr}. Yet, in spite of this, an elaborate study combining elements of QCD, large $N_c$ limits and phenomenology concludes that while low lying scalar mesons, like the $\sigma$, $f_0(980)$, $\cdots$ are completely off the $q\bar q$ picture, the vector mesons are largely $q\bar q$ states \cite{Pelaez:2015qba}. Our result comes handy when some calculations could make us lose confidence in this picture. Indeed, in Ref.~\cite{Barnes:2007xu}, where a calculation within a quark model was done to assess the relevance of the meson-meson components in the vector mesons, even the $J/\psi$ was found to have a $Z$ probability of only $65\%$, implying that more massive $\psi$ vectors could have an even smaller $Z$ probability. The result of the present paper incorporating the features of the $\psi(3770)$ shape in the $e^+e^-\to D\bar D$ reactions, demanded the presence of a form factor that has a consequence the small meson-meson probabilities and the large $Z$ value.

\section{Summary and discussion}\label{conclusion}
We have performed an evaluation of the meson-meson components in the $\psi(3770)$ wave function, considering $PP$, $PV$, $VP$ and $VV$ components. We found that the determination of such probabilities was much tied to the shape of the $e^+e^-\to D\bar D$ reaction, which we described in terms of the $\psi(3770)$ selfenergy due to the meson-meson components. Indeed, the shape of the cross section for this reaction determined the range of a form factor that was determined in the evaluation of the meson-meson probabilities of the $\psi(3770)$ wave function. Within uncertainties we found that the $Z$ probability of a vector component in the $\psi(3770)$ is of the order of $80\sim85\%$ and the individual meson-meson components are small. This finding is very important, extracting from this phenomenological study the same conclusion obtained from QCD and large $N_c$ behavior, plus meson-meson scattering data, that vector mesons are largely $q\bar q$ objects \cite{Pelaez:2015qba}. This is also in line with $Z$ evaluation for the $\rho$ with a different method which gives $Z\sim0.75$, even with such a large width for the decay to two pions \cite{Aceti:2012dd}.

\section{ACKNOWLEDGEMENT}
Q.~X.~Yu acknowledges the support from the National Natural Science Foundation of China (Grant No.~11775024 and 11575023). W.~H.~Liang acknowledges the support from the National Natural Science Foundation of China (Grant No.~11565007 and 11847317). This work is partly supported by the Spanish Ministerio de Economia y Competitividad and European FEDER funds under Contracts No. FIS2017-84038-C2-1-P B and No. FIS2017-84038-C2-2-P B, and the
Generalitat Valenciana in the program Prometeo II-2014/068, and the project Severo Ochoa
of IFIC, SEV-2014-0398.

\begin{appendix}
\section{Evaluation of the $\psi(3770)$ coupling to $D\bar D,D\bar D^\ast,D^\ast\bar D,D^\ast\bar D^\ast$}\label{appa}
According to Ref.~\cite{Godfrey:1985xj} the $\psi(3770)$ is a $1 ^3D_1$ state. This means radial wave function in the ground state, spin $1$ and angular momentum of the two quarks $L=2$, coupling later $L=2$ with $S=1$ to give $J=1$. We start with the $c\bar c$ spin wave function
\begin{align}\label{spin1}
|S\tilde M^\prime\rangle=|1\tilde M^\prime\rangle&=\sum_{m,m^\prime}\mathcal C(s_1,s_2,S;m,m^\prime,\tilde M^\prime)|s_1,m\rangle|s_2,m^\prime\rangle\nonumber\\
&=\sum_{m,m^\prime}\mathcal C(\frac{1}{2},\frac{1}{2},1;m,m^\prime,\tilde M^\prime)|\frac{1}{2},m\rangle|\frac{1}{2},m^\prime\rangle,
\end{align}
where $s_1$ and $s_2$ correspond to the spin of $c$ and $\bar c$ in Fig.~\ref{feyn1}, and $m$, $m^\prime$ are their third components respectively, while $S$ and $\tilde M^\prime$ are the total spin and third component of $c\bar c$. Then after coupling the spin part to the orbital part of $c\bar c$, we have
\begin{align}\label{spin2}
|J\tilde M\rangle=|1\tilde M\rangle&=\sum_{M_3^\prime,\tilde M^\prime}\mathcal C(L,S,J;M_3^\prime,\tilde M^\prime,\tilde M)Y_{L,M_3^\prime}(\boldsymbol{\hat r})|S,\tilde M^\prime\rangle\nonumber\\
&=\sum_{M_3^\prime,\tilde M^\prime}\mathcal C(2,1,1;M_3^\prime,\tilde M^\prime,\tilde M)Y_{2,M_3^\prime}(\boldsymbol{\hat r})|1,\tilde M^\prime\rangle.
\end{align}
We do the same to couple the spin and orbital angular momentum of the $q\bar q$ vacuum state $^{3}P_0$ in Fig.~\ref{feyn1}, as done in Refs.~\cite{Dai:2018thd,Liang:2016ydj},
\begin{equation}\label{spin3}
|1S_3\rangle=\sum_{s}\mathcal C(\frac{1}{2},\frac{1}{2},1;s,S_3-s)|\frac{1}{2},s\rangle|\frac{1}{2},S_3-s\rangle,
\end{equation}
and we combine this state, $|1S_3\rangle$, with the $L=1$ state $Y_{1,M_3}(\boldsymbol{\hat r})$ to give $J=0$,
\begin{equation}\label{spin4}
|00\rangle=\sum_{M_3}\mathcal C(1,1,0;M_3,S_3)Y_{1,M_3}(\boldsymbol{\hat r})|1,S_3\rangle,
\end{equation}
implying $M_3+S_3=0$, $i.e.$, $M_3=-S_3$, which allows us to rewrite Eq.~\eqref{spin4} as follows
\begin{align}\label{spin41}
|00\rangle&=\sum_{S_3}\mathcal C(1,1,0;-S_3,S_3)Y_{1,-S_3}(\boldsymbol{\hat r})|1,S_3\rangle\nonumber\\
&=\sum_{S_3}(-1)^{1+S_3}\frac{1}{\sqrt 3}Y_{1,-S_3}(\boldsymbol{\hat q})|1,S_3\rangle.
\end{align}

In addition we have the spatial matrix element, where the $c,\bar c$ quark states are in their ground state. Then we have
\begin{equation}\label{spatial}
ME(\boldsymbol q)=\int d^3\boldsymbol r\varphi_c(r)\varphi_q(r)\varphi_{\bar q}(r)\varphi_{\bar c}(r)e^{i\boldsymbol{q\cdot r}}Y_{1,-S_3}(\boldsymbol{\hat r})Y_{2,M_3^\prime}(\boldsymbol{\hat r}),
\end{equation}
where $\boldsymbol q$ is the exchanged momentum between the two mesons produced after the hadronization, and $e^{i\boldsymbol{q\cdot r}}$ can be expanded as
\begin{equation}\label{exponential}
e^{i\boldsymbol{q\cdot r}}=4\pi\sum_l i^lj_l(qr)Y_{l\mu}(\boldsymbol{\hat q})Y_{l\mu}^\ast(\boldsymbol{\hat r}).
\end{equation}
The coupling rule for spherical harmonics permits an easy way of combining three spherical harmonic functions as we show in the following equation, where two of them come from Eq.~\eqref{spatial} and the other one, $Y_{l,\mu}^\ast(\boldsymbol{\hat r})$, from Eq.~\eqref{exponential}. After integrating over the full solid angle, we arrive at \cite{Rose:1957}
\begin{equation}\label{combine1}
\int d\Omega\,Y^\ast_{l\mu}(\boldsymbol{\hat r})Y_{1,-S_3}(\boldsymbol{\hat r})Y_{2,M_3^\prime}(\boldsymbol{\hat r})=\left(\frac{15}{4\pi(2l+1)}\right)^{1/2}\mathcal C(2,1,l;M_3^\prime,-S_3,\mu)\mathcal C(2,1,l;0,0,0),
\end{equation}
where for parity reasons $2+1+l$ must be even, hence, $l=1,3$, but $l=1$ is required to have a $P$-wave coupling of $J/\psi$ to $D\bar D$ at the end, such that we obtain (where we use $\mathcal C(2,1,1;0,0,0)=-\sqrt\frac{2}{5}$)
\begin{equation}\label{me1}
ME(\boldsymbol q)=-4\pi i\,Y_{1,M^\prime_3-S_3}(\boldsymbol{\hat q})\sqrt\frac{2}{4\pi}\,\mathcal C(2,1,1;M_3^\prime,-S_3,M_3^\prime-S_3)\int r^2dr\varphi_c(r)\varphi_q(r)\varphi_{\bar q}(r)\varphi_{\bar c}(r)j_1(qr).
\end{equation}
Since $j_1(qr)$ goes as $qr$ for small values of $qr$, $ME(\boldsymbol q)$ grows linearly $q$ for small $q$, and for that reason we rewrite $ME(\boldsymbol q)$ as
\begin{equation}\label{me2}
ME(\boldsymbol q)=-\frac{4\pi i}{3}q\,Y_{1,M^\prime_3-S_3}(\boldsymbol{\hat q})\sqrt\frac{2}{4\pi}\,\mathcal C(2,1,1;M_3^\prime,-S_3,M_3^\prime-S_3)\int r^2dr\prod_i\varphi_i(r)\frac{3j_1(qr)}{qr}r,
\end{equation}
where the factor $\frac{3j_1(qr)}{qr}$ goes to $1$ as $qr$ approaches 0 and is a smooth function, such that the integral in Eq.~\eqref{me2} is a smooth function of $q$ for small $q$, the typical form of the form factors and the form that we will take for our empirical form factors. We can write $qY_{1,M_3^\prime-S_3}(\boldsymbol{\hat q})$ in Eq.~\eqref{me2} as $\sqrt\frac{3}{4\pi}q_{M_3^\prime-S_3}$ (in spherical basis), which accounts for the vector coupling to two pseudoscalars.

At the same time, by coupling the vacuum state $|00\rangle$ with $c$, $\bar c$ spins we can obtain the final angular momenta of the two mesons produced, $|J_1M_2\rangle$ and $|J_2 M_2\rangle$, which is accomplished by means of the Clebsch-Gordan coefficients,
\begin{equation}\label{j1}
|J_1M_1\rangle=\sum_m\mathcal C(\frac{1}{2},\frac{1}{2},J_1;m,s,M_1)|\frac{1}{2},m\rangle|\frac{1}{2},s\rangle,
\end{equation}
\begin{equation}\label{j2}
|J_2M_2\rangle=\sum_{m^\prime}\mathcal C(\frac{1}{2},\frac{1}{2},J_2;S_3-s,m^\prime,M_2)|\frac{1}{2},S_3-s\rangle|\frac{1}{2},m^\prime\rangle,
\end{equation}
where we obtain the constrains: $m+s=M_1$, $S_3-s+m^\prime=M_2$, leading to $m=M_1-s$, $m^\prime=M_2-S_3+s$. Further constrains between $S_3$ and $M_1$, $M_2$ can be derived with the help of Eq.~\eqref{spin1}, and $S_3$ satisfies the relation, $S_3=M_1+M_2-\tilde M^\prime$.

Finally, we can write down the matrix element of the transition from $|1\tilde M^\prime\rangle$ to $|J_1M_1\rangle|J_2M_2\rangle$ by combining Eqs.~\eqref{spin1}, \eqref{spin2}, \eqref{spin3}, \eqref{spin41}, \eqref{j1} and \eqref{j2},
\begin{align}\label{trans}
ME=&-\frac{4\pi i}{3}\sqrt\frac{2}{4\pi}\sum_{\tilde M^\prime}\sum_s\sum_{S_3}\mathcal C(2,1,1;M_3^\prime,-S_3,M_3^\prime-S_3)\,\mathcal C(2,1,1;M_3^\prime,\tilde M^\prime,\tilde M)q\,Y_{1,M^\prime_3-S_3}(\boldsymbol{\hat q})\nonumber\\
&\times\mathcal C(\frac{1}{2},\frac{1}{2},1;M_1-s,M_2-S_3+s,\tilde M^\prime)\,\mathcal C(\frac{1}{2},\frac{1}{2},1;s,S_3-s,S_3)(-1)^{1+S_3}\frac{1}{\sqrt 3}\nonumber\\
&\times\mathcal C(\frac{1}{2},\frac{1}{2},J_1;M_1-s,s,M_1)\,\mathcal C(\frac{1}{2},\frac{1}{2},J_2;S_3-s,M_2-S_3+s,M_2).
\end{align}
Now we use $S_3=M_1+M_2-\tilde M^\prime$ and the above equation can be rewritten as,
\begin{align}\label{trans2}
ME=&-\frac{4\pi i}{3}\sqrt\frac{2}{4\pi}\sum_s\sum_{\tilde M^\prime}\mathcal C(2,1,1;\tilde M-\tilde M^\prime,\tilde M^\prime-M_1-M_2,\tilde M-M_1-M_2)\nonumber\\
&\times\mathcal C(2,1,1;\tilde M-\tilde M^\prime,\tilde M^\prime,\tilde M)q\,Y_{1,\tilde M-M_1-M_2}(\boldsymbol{\hat q})(-1)^{1+M_1+M_2-\tilde M^\prime}\frac{1}{\sqrt 3}\nonumber\\
&\times\mathcal C(\frac{1}{2},\frac{1}{2},1;M_1-s,\tilde M^\prime-M_1+s,\tilde M^\prime)\,\mathcal C(\frac{1}{2},\frac{1}{2},1;s,M_1+M_2-\tilde M^\prime-s,M_1+M_2-\tilde M^\prime)\nonumber\\
&\times\mathcal C(\frac{1}{2},\frac{1}{2},J_1;M_1-s,s,M_1)\,\mathcal C(\frac{1}{2},\frac{1}{2},J_2;M_1+M_2-\tilde M^\prime-s,\tilde M^\prime-M_1+s,M_2),
\end{align}

In Eq.~\eqref{trans2} there are four CG coefficients that depend on $s$. In order to get an expression with three CG coefficients to be written in terms of Racah coefficients we proceed as follows. Firstly, we need to permute some indices in the fourth CG coefficient in Eq.~\eqref{trans2} as Ref.~\cite{Rose:1957},
\begin{align}\label{per1}
&\mathcal C(\frac{1}{2},\frac{1}{2},1;s,M_1+M_2-\tilde M^\prime-s,M_1+M_2-\tilde M^\prime)\nonumber\\
=&(-1)^{1/2-s}\sqrt\frac{3}{2}\,\mathcal C(1,\frac{1}{2},\frac{1}{2};M_1+M_2-\tilde M^\prime,-s,M_1+M_2-\tilde M^\prime-s),
\end{align}
and together with the last one in Eq.~\eqref{trans2}, we can convert them into other two CG coefficients where only one CG coefficient depends on $s$ \cite{Rose:1957},
\begin{align}\label{per2}
&\mathcal C(1,\frac{1}{2},\frac{1}{2};M_1+M_2-\tilde M^\prime,-s,M_1+M_2-\tilde M^\prime-s)\,\mathcal C(\frac{1}{2},\frac{1}{2},J_2;M_1+M_2-\tilde M^\prime-s,\tilde M^\prime-M_1+s,M_2) \nonumber\\
=&\sum_{j^{''}}\sqrt{2(2j^{''}+1)}\,\mathcal W(1,\frac{1}{2},J_2,\frac{1}{2};\frac{1}{2},j^{''})\,\mathcal C(\frac{1}{2},\frac{1}{2},j^{''};-s,-M_1+\tilde M^\prime+s,-M_1+\tilde M^\prime)\nonumber\\
&\times\mathcal C(1,j^{''},J_2;M_1+M_2-\tilde M^\prime,-M_1+\tilde M^\prime,M_2),
\end{align}
where $\mathcal W$ is a Racah coefficient \cite{Rose:1957}. Similarly, we need to permute indices of the third CG coefficient in Eq.~\eqref{trans2} and the first CG coefficient in Eq.~\eqref{per2} before we move on to the next combination,
\begin{align}\label{per3}
&\mathcal C(\frac{1}{2},\frac{1}{2},1;M_1-s,\tilde M^\prime-M_1+s,\tilde M^\prime)\nonumber\\
=&(-1)^{1+1/2-M_1+\tilde M^\prime+s}\sqrt\frac{3}{2}\,\mathcal C(1,\frac{1}{2},\frac{1}{2};\tilde M^\prime,M_1-\tilde M^\prime-s,M_1-s),
\end{align}
and
\begin{align}\label{per4}
&\mathcal C(\frac{1}{2},\frac{1}{2},j^{''};-s,-M_1+\tilde M^\prime+s,-M_1+\tilde M^\prime)\nonumber\\
=&[(-1)^{1/2+1/2-j^{''}}]^2\,\mathcal C(\frac{1}{2},\frac{1}{2},j^{''};M_1-\tilde M^\prime-s,s,M_1-\tilde M^\prime).
\end{align}

We combine now the three CG coefficients from Eqs.~\eqref{per3}, \eqref{per4} and the fifth CG coefficient in Eq.~\eqref{trans2} \cite{Rose:1957}, and since the phase does not depend on $s$, we can write
\begin{align}\label{per5}
&\sum_s\mathcal C(1,\frac{1}{2},\frac{1}{2};\tilde M^\prime,M_1-\tilde M^\prime-s,M_1-s)\,\mathcal C(\frac{1}{2},\frac{1}{2},J_1;M_1-s,s,M_1)\nonumber\\
&\times \mathcal C(\frac{1}{2},\frac{1}{2},j^{''};M_1-\tilde M^\prime-s,s,M_1-\tilde M^\prime)\nonumber\\
=&\sqrt{2(2j^{''}+1)}\,\mathcal W(1,\frac{1}{2},J_1,\frac{1}{2};\frac{1}{2},j^{''})\,\mathcal C(1,j^{''},J_1;\tilde M^\prime,M_1-\tilde M^\prime,M_1),
\end{align}
such that Eq.~\eqref{trans2} can be rewritten as
\begin{align}\label{trans4}
ME=&-\frac{4\pi i}{3}q\,Y_{1,\tilde M-M_1-M_2}(\boldsymbol{\hat q})\sqrt\frac{2}{4\pi}\sum_{\tilde M^\prime}\sum_{j^{''}}\big[\sqrt 3\,(-1)^{1+M_2}(2j^{''}+1)\big]\prod_{i=1}^4\mathcal C_i\prod_{j=1}^2\mathcal W_j.
\end{align}
where $\prod_{i=1}^4\mathcal C_i\prod_{j=1}^2\mathcal W_j$ can be expressed explicitly as follows,
\begin{align}\label{cw}
\prod_{i=1}^4\mathcal C_i\prod_{j=1}^2\mathcal W_j=\,&\mathcal C(2,1,1;\tilde M-\tilde M^\prime,\tilde M^\prime-M_1-M_2,\tilde M-M_1-M_2)\,\mathcal C(2,1,1;\tilde M-\tilde M^\prime,\tilde M^\prime,\tilde M)\nonumber\\
\times\,&\mathcal C(1,j^{''},J_2;M_1+M_2-\tilde M^\prime,-M_1+\tilde M^\prime,M_2)\,\mathcal C(1,j^{''},J_1;\tilde M^\prime,M_1-\tilde M^\prime,M_1)\nonumber\\
\times\,&\mathcal W(1,\frac{1}{2},J_1,\frac{1}{2};\frac{1}{2},j^{''})\,\mathcal W(1,\frac{1}{2},J_2,\frac{1}{2};\frac{1}{2},j^{''}).
\end{align}

Next we begin evaluating different cases with $J_1$ and $J_2$ assigned to particular values, we start with the case where $J_1=0$, $J_2=0$, which corresponds to the $PP$ coupling,
\begin{enumerate}
\item[(i)] $PP$: $J_1=0$, $J_2=0$\\
It implies $M_1=0$, $M_2=0$, and Eq.~\eqref{per2} leads us to fact that $j^{''}$ can only be $1$ in this case. With these particular quantum numbers we can easily obtain the Racah coefficients,
\begin{equation}
\mathcal W(1,\frac{1}{2},0,\frac{1}{2};\frac{1}{2},1)=\frac{1}{\sqrt 6},
\end{equation}
and two of the CG coefficients
\begin{eqnarray}
\mathcal C(1,1,0;-\tilde M^\prime,\tilde M^\prime,0)&=&(-1)^{1+\tilde M^\prime}\sqrt\frac{1}{3}\\
\mathcal C(1,1,0;\tilde M^\prime,-\tilde M^\prime,0)&=&(-1)^{1-\tilde M^\prime}\sqrt\frac{1}{3}.
\end{eqnarray}
Permuting the first two indices in the first two CG coefficients of Eq.~\eqref{cw} we obtain the following equation for $|ME|^2$ in this case
\begin{align}\label{pp1}
\Big|-\frac{4\pi i}{3}\Big|^2q^2Y_{1,\tilde M}(\boldsymbol{\hat q})Y_{1,\tilde M}^\ast(\boldsymbol{\hat q})\frac{2}{4\pi}[\sqrt 3\times 3]^2\Big[\sum_{\tilde M^\prime}\mathcal C(1,2,1;\tilde M^\prime,\tilde M-\tilde M^\prime,\tilde M)^2\Big]^2(\frac{1}{3})^2(\frac{1}{6})^2,
\end{align}
further simplification can be done by replacing $Y_{1,\tilde M}(\boldsymbol{\hat q})Y_{1,\tilde M}^\ast(\boldsymbol{\hat q})$ with \begin{align}\label{yy}
\frac{1}{4\pi}\int d\Omega\,Y_{1,\tilde M}(\boldsymbol{\hat q})Y_{1,\tilde M}^\ast(\boldsymbol{\hat q})=\frac{1}{4\pi},
\end{align}
as we have to integrate over angles in $\int d^3q$ of the loop. Since
\begin{align}
\sum_{\tilde M^\prime}\,\mathcal C(1,2,1;\tilde M^\prime,\tilde M-\tilde M^\prime,\tilde M)^2=1,
\end{align}
we next sum and average $|ME|^2$ over $\tilde M$ and we arrive at the final result for $|ME|^2$ summed and averaged over $\tilde M$ of Eq.~\eqref{pp1}, which is
\begin{align}
\overline{\sum_{\tilde M}}|ME|^2=\frac{1}{12}\Big|\frac{4\pi i}{3}\Big|^2q^2\frac{2}{4\pi}\frac{1}{4\pi}.
\end{align}
\item[(ii)]$PV$: $J_1=1$, $J_2=0$\\
In this case, we have $M_2=0$, and $j^{''}$ can be determined with the constrains in Eq.~\eqref{cw}, hence, since $1+j^{''}$ must give $J_2=0$, $j^{''}=1$. Similarly, we can obtain the Racah coefficients in Eq.~\eqref{trans4} with these specific quantum numbers,
\begin{eqnarray}
\mathcal W_1(1,\frac{1}{2},1,\frac{1}{2};\frac{1}{2},1)&=&\frac{1}{3},\\
\mathcal W_2(1,\frac{1}{2},0,\frac{1}{2};\frac{1}{2},1)&=&\frac{1}{\sqrt 6},
\end{eqnarray}
one of the CG coefficients in Eq.~\eqref{trans4},
\begin{align}
\mathcal C(1,1,0;M_1-\tilde M^\prime,-M_1+\tilde M^\prime,0)=(-1)^{1-M_1+\tilde M^\prime}\sqrt\frac{1}{3},
\end{align}
and the other three CG coefficients can be combined together to give
\begin{align}
&\sum_{\tilde M^\prime}\mathcal C(2,1,1;\tilde M-\tilde M^\prime,\tilde M^\prime-M_1,\tilde M-M_1)\,\mathcal C(2,1,1;\tilde M-\tilde M^\prime,\tilde M^\prime,\tilde M)\nonumber\\
&\times\mathcal C(1,1,1;\tilde M^\prime,M_1-\tilde M^\prime,M_1)(-1)^{1-M_1+\tilde M^\prime}\nonumber\\
=&\sum_{\tilde M^\prime}(-1)^{1+\tilde M^\prime}\sqrt\frac{3}{5}\,\mathcal C(1,1,2;\tilde M,-\tilde M^\prime,\tilde M-\tilde M^\prime)\,\mathcal C(2,1,1;\tilde M-\tilde M^\prime,\tilde M^\prime-M_1,\tilde M-M_1)\nonumber\\
&\times(-1)\mathcal C(1,1,1;-\tilde M^\prime,\tilde M^\prime-M_1,-M_1)(-1)^{1-M_1+\tilde M^\prime}\nonumber\\
=&(-1)^{1-M_1}\sqrt\frac{3}{5}\sqrt {15}\,\mathcal W(1,1,1,1;2,1)\,\mathcal C(1,1,1;\tilde M,-M_1),
\end{align}
where
\begin{align}
\mathcal W(1,1,1,1;2,1)=\frac{1}{6}.
\end{align}
Then, we have a similar equation for $|ME|^2$ in this case after multiplying all terms and squaring, we get for $\overline\sum_{\tilde M}\sum_{M_1}|ME|^2$
\begin{align}
&\Big|-\frac{4\pi i}{3}\Big|^2q^2Y_{1,\tilde M-M_1}(\boldsymbol{\hat q})Y_{1,\tilde M-M_1}^\ast(\boldsymbol{\hat q})\frac{2}{4\pi}[\sqrt 3\times3]^2\nonumber\\
\times&\frac{1}{3}\Big[\sum_{M_1}\sum_{\tilde M}\mathcal C^2(1,1,1;\tilde M,M_1-\tilde M,M_1)\Big]^2(\frac{3}{5})(15)(\frac{1}{6})^2(\frac{1}{3})(\frac{1}{3})^2(\frac{1}{6}),
\end{align}
and using the equivalent equation to Eq.~\eqref{yy} for the spherical harmonics, we have 
\begin{align}
\overline{\sum_{\tilde M}}\sum_{M_1}|ME|^2=\frac{1}{24}\Big|\frac{4\pi i}{3}\Big|^2q^2\frac{2}{4\pi}\frac{1}{4\pi}.
\end{align}
\item[(iii)]$VP$: $J_1=0$, $J_2=1$\\
We follow closely the previous case (ii) and obtain the same result for $|ME|^2$ in this scenario,
\begin{align}
\overline{\sum_{\tilde M}}\sum_{M_2}|ME|^2=\frac{1}{24}\Big|\frac{4\pi i}{3}\Big|^2q^2\frac{2}{4\pi}\frac{1}{4\pi}.
\end{align}
\item[(iv)]$VV$: $J_1=1$, $J_2=1$\\
The calculations in this case is relatively complicated since $j^{''}$ now can be both $0$ and $1$ (see CG coefficient in Eq.~\eqref{per2}). We thus separate these two situations and present the case with $j^{''}=0$ first.
\begin{enumerate}
\item[(a)]$j^{''}=0$\\
As always, first we have the two Racah coefficients of Eq.~\eqref{trans4}, which are the same in this case
\begin{align}
\mathcal W(1,\frac{1}{2},1,\frac{1}{2};\frac{1}{2},0)=-\frac{1}{\sqrt 6},
\end{align}
as for the two of the CG coefficients in Eq.~\eqref{trans4} that contain $j^{''}$, we have
\begin{eqnarray}
\mathcal C(1,0,1;M_1+M_2-\tilde M^\prime,-M_1+\tilde M^\prime,M_2)&=&1,\\
\mathcal C(1,0,1;\tilde M^\prime,M_1-\tilde M^\prime,M_1)&=&1,
\end{eqnarray}
with the condition that $\tilde M^\prime=M_1$. Furthermore, the other two CG coefficients in Eq.~\eqref{trans4} can be rewritten as
\begin{align}\label{ran1}
&\sum_{\tilde M^\prime}\mathcal C(2,1,1;\tilde M-\tilde M^\prime,\tilde M^\prime-M_1-M_2,\tilde M-M_1-M_2)\,\mathcal C(2,1,1;\tilde M-\tilde M^\prime,\tilde M^\prime,\tilde M)\delta_{\tilde M^\prime,M_1}\nonumber\\
=&\mathcal C(2,1,1;\tilde M-M_1,-M_2,\tilde M-M_1-M_2)\,\mathcal C(2,1,1;\tilde M-M_1,M_1,\tilde M),
\end{align}
the square of the Eq.~\eqref{ran1} gives us
\begin{align}\label{ran2}
&\mathcal C(2,1,1;\tilde M-M_1,-M_2,\tilde M-M_1-M_2)^2\,\mathcal C(2,1,1;\tilde M-M_1,M_1,\tilde M)^2\nonumber\\
=&\frac{3}{5}\,\mathcal C(1,1,2;M_2,\tilde M-M_1-M_2,\tilde M-M_1)^2\,\mathcal C(2,1,1;\tilde M-M_1,M_1,\tilde M)^2,
\end{align}
where we write the second term $\mathcal C(2,1,1;\tilde M-M_1,M_1,\tilde M)^2$ as $\mathcal C(2,1,1;\tilde M-M_1,\tilde M-(\tilde M-M_1),\tilde M)^2$, and then sum over $M_2$, $\tilde M-M_1$ and $\tilde M$. We obtain the following values with $\tilde M-M_1$ and $\tilde M$ fixed,
\begin{align}
\sum_{M_2}\mathcal C(1,1,2;M_2,\tilde M-M_1-M_2,\tilde M-M_1)^2=1,
\end{align}
and with $\tilde M$ fixed, we have
\begin{align}
\sum_{\tilde M-M_1}\mathcal C(2,1,1;\tilde M-M_1,\tilde M-(\tilde M-M_1),\tilde M)^2=1,
\end{align}
and the sum over $\tilde M$ gives $3$. We shall take the factor $\frac{1}{3}$ from the average at the end.

Finally, following the same steps used in the previous cases, we obtain $|ME|^2$ in this case,
\begin{align}
\sum_{\tilde M}\sum_{M_1}\sum_{M_2}|ME|_a^2=\frac{3}{20}\Big|\frac{4\pi i}{3}\Big|^2q^2\frac{2}{4\pi}\frac{1}{4\pi}.
\end{align}
\item[(b)]$j^{''}=1$\\
In this case, we have two of the CG coefficients in Eq.~\eqref{cw} that can be rewritten as
\begin{align}\label{cc1}
&\mathcal C(1,j^{''},J_2;M_1+M_2-\tilde M^\prime,-M_1+\tilde M^\prime,M_2)\,\mathcal C(1,j^{''},J_1;\tilde M^\prime,M_1-\tilde M^\prime,M_1)\nonumber\\
=&\mathcal C(1,1,1;M_1+M_2-\tilde M^\prime,-M_1+\tilde M^\prime,M_2)\,\mathcal C(1,1,1;\tilde M^\prime,M_1-\tilde M^\prime,M_1)\nonumber\\
=&(-1)^{-M_1-M_2}\mathcal C(1,1,1;M_2,\tilde M^\prime-M_1-M_2,\tilde M^\prime-M_1)\,\mathcal C(1,1,1;M_1,-\tilde M^\prime,M_1-\tilde M^\prime)\nonumber\\
=&(-1)^{1+\tilde M^\prime}\mathcal C(1,1,1;M_1,-\tilde M^\prime,M_1-\tilde M^\prime)\,\mathcal C(1,1,1;-\tilde M^\prime+M_1,\tilde M^\prime-M_1-M_2,-M_2)\nonumber\\
=&\sum_{j^{'''}}(-1)^{1+\tilde M^\prime}[3(2j^{'''}+1)]^{1/2}\,\mathcal W(1,1,1,1;1,j^{'''})\nonumber\\
&\times\mathcal C(1,1,j^{'''};-\tilde M^\prime,\tilde M^\prime-M_1-M_2)\mathcal C(1,j^{'''},1;M_1,-M_1-M_2),
\end{align}
where we separate the CG coefficients and only one depends on $\tilde M^\prime$, and that one can be combined together with the other two CG coefficients of Eq.~\eqref{cw} to give
\begin{align}
&\sum_{\tilde M^\prime}(-1)^{1+\tilde M^\prime}\mathcal C(2,1,1;\tilde M-\tilde M^\prime,\tilde M^\prime-M_1-M_2,\tilde M-M_1-M_2)\nonumber\\
&\times\mathcal C(2,1,1;\tilde M-\tilde M^\prime,\tilde M^\prime,\tilde M)\,\mathcal C(1,1,j^{'''};-\tilde M^\prime,\tilde M^\prime-M_1-M_2)\nonumber\\
=&\mathcal C(1,1,2;\tilde M,-\tilde M^\prime,\tilde M-\tilde M^\prime)\,\mathcal C(2,1,1;\tilde M-\tilde M^\prime,\tilde M^\prime-M_1-M_2,\tilde M-M_1-M_2)\nonumber\\
&\times\mathcal C(1,1,j^{'''};-\tilde M^\prime,\tilde M^\prime-M_1-M_2)\nonumber\\
=&[5(2j^{'''}+1)]^{1/2}\,\mathcal W(1,1,1,1;2,j^{'''})\,\mathcal C(1,j^{'''},1;\tilde M,-M_1-M_2).
\end{align}
In this way, we now have the following equation for Eq.~\eqref{cw} in this case
\begin{align}\label{cc2}
&\sum_{j^{'''}}\frac{\sqrt{15}}{9}(2j^{'''}+1)\,\mathcal W(1,1,1,1;1,j^{'''})\,\mathcal W(1,1,1,1;2,j^{'''})\nonumber\\
&\times\mathcal C(1,j^{'''},1;M_1,-M_1-M_2)\,\mathcal C(1,j^{'''},1;\tilde M,-M_1-M_2)\nonumber\\
=&\sum_{j^{'''}}(-1)^{-M_1-\tilde M}(\frac{\sqrt {15}}{3})\,\mathcal W(1,1,1,1;1,j^{'''})\,\mathcal W(1,1,1,1;2,j^{'''})\nonumber\\
&\times\mathcal C(1,1,j^{'''};M_1,M_2,M_1+M_2)\,\mathcal C(1,1,j^{'''};\tilde M,M_1+M_2-\tilde M,M_1+M_2),
\end{align}
similarly, for these two CG coefficients in Eq.~\eqref{cc2} we will sum over $M_1$, $\tilde M$, $M_1+M_2$ when we square, which leads us to
\begin{align}
\sum_{M_1}\mathcal C(1,1,j^{'''}_1;M_1,M_2,M_1+M_2)\,\mathcal C(1,1,j^{'''}_2;M_1,M_2,M_1+M_2)=\delta_{j^{'''}_1,\,j^{'''}_2},
\end{align}
where we keep $\tilde M$ and $M_1+M_2$ fixed, and a similar thing can be done to the other CG coefficients when we square
\begin{align}
&\sum_{\tilde M}\mathcal C(1,1,j^{'''}_1;\tilde M,M_1+M_2-\tilde M,M_1+M_2)\,\mathcal C(1,1,j^{'''}_2;\tilde M,M_1+M_2-\tilde M,M_1+M_2)\nonumber\\
&=\delta_{j^{'''}_1,\,j^{'''}_2},
\end{align}
and sum over $M_1+M_2$ will give us a factor of $3$, which is the same as we obtained in the last case. Then we have the following equation when we square Eq.~\eqref{cc2}
\begin{align}\label{wwww}
&\sum_{j^{'''}}(3\times\frac{15}{9})\,\mathcal W(1,1,1,1;1,j^{'''})^2\,\mathcal W(1,1,1,1;2,j^{'''})^2\nonumber\\
=&\sum_{j^{'''}}5\,\mathcal W(1,1,1,j^{'''};1,1)^2\,\mathcal W(2,1,1,j^{'''};1,1)^2\nonumber\\
=&\frac{1}{81}\frac{426}{80},
\end{align}
where we sum over $j^{'''}$ for $j^{'''}=0,1,2$ and all the Racah coefficients used in Eq.~\eqref{wwww} are listed below. 
\begin{align}
  &\mathcal W(1,1,1,0;1,1)=\frac{1}{3},\,\,\,\,\,\,\,\,\,\mathcal W(2,1,1,0;1,1)=\frac{1}{3},\nonumber\\
  &\mathcal W(1,1,1,1;1,1)=\frac{1}{6},\,\,\,\,\,\,\,\,\,\mathcal W(2,1,1,1;1,1)=-\frac{1}{6},\nonumber\\
  &\mathcal W(1,1,1,2;1,1)=-\frac{1}{6},\,\,\,\,\,\mathcal W(2,1,1,2;1,1)=\frac{1}{30}.
\end{align}
Consequently, we have $|ME|^2$ in this case as
\begin{align}
\sum_{\tilde M}\sum_{M_1}\sum_{M_2}|ME|_b^2&=27\times\frac{1}{81}\frac{426}{80}\Big|\frac{4\pi i}{3}\Big|^2q^2\frac{2}{4\pi}\frac{1}{4\pi}\nonumber\\
&=\frac{213}{120}\Big|\frac{4\pi i}{3}\Big|^2q^2\frac{2}{4\pi}\frac{1}{4\pi}.
\end{align}
\end{enumerate}

Crossed terms are calculated to be $0$ in this particular case, and we then add up parts (a) and (b) taking into account the factor $(\frac{1}{3})$ from the average over $\tilde M$. Then we arrive at
\begin{align}
\overline{\sum_{\tilde M}}\sum_{M_1}\sum_{M_2}(|ME|_a^2+|ME|_b^2)=\frac{231}{360}\Big|\frac{4\pi i}{3}\Big|^2q^2\frac{2}{4\pi}\frac{1}{4\pi}.
\end{align}
\end{enumerate}

To sum it up, we have obtained all the scattering amplitudes $\overline\sum\sum |t|^2$ with different types of interactions: $PP$, $PV$, $VP$ and $VV$, we present here again for clarity
\begin{eqnarray}\label{tt}
PP &:& \quad\quad\quad\frac{1}{12}\Big|\frac{4\pi i}{3}\Big|^2q^2\frac{2}{4\pi}\frac{1}{4\pi},\nonumber\\
PV &:& \quad\quad\quad\frac{1}{24}\Big|\frac{4\pi i}{3}\Big|^2q^2\frac{2}{4\pi}\frac{1}{4\pi},\nonumber\\
VP &:& \quad\quad\quad\frac{1}{24}\Big|\frac{4\pi i}{3}\Big|^2q^2\frac{2}{4\pi}\frac{1}{4\pi},\nonumber\\
VV &:& \quad\quad\quad\frac{231}{360}\Big|\frac{4\pi i}{3}\Big|^2q^2\frac{2}{4\pi}\frac{1}{4\pi}.
\end{eqnarray}
On top of that, there is a constant common to all the decay modes which would appear in the hadronization process. Then we can omit $|-\frac{4\pi i}{3}|^2\frac{1}{4\pi}\frac{2}{4\pi}$ in Eq.~\eqref{tt} and replace it with a factor $|A|^2$, which is fitted to the experimental data.
\end{appendix}

\end{document}